\documentclass[journal]{IEEEtran}
\usepackage{cite,amssymb}
\usepackage[pdftex]{graphicx}
\usepackage{algorithm}
\usepackage[cmex10]{amsmath}
\usepackage{algorithm}
\usepackage{algorithm}
\usepackage{algpseudocode}
\usepackage{array,balance}

\ifCLASSOPTIONcompsoc
\usepackage[caption=false,font=normalsize,labelfont=sf,textfont=sf]{subfig}
\else
\usepackage[caption=false,font=footnotesize]{subfig}
\fi
\usepackage{url}
\usepackage{amsfonts}
\usepackage{exscale}
\usepackage{relsize}
\usepackage{multicol,color}
\newcommand{\mH}{\mathop{\rm H}}
\newcommand{\mT}{\mathop{\rm T}}
\begin{document}
\title{Pattern Synthesis via Complex-Coefficient Weight Vector Orthogonal Decomposition--Part II:\\ Robust Sidelobe Synthesis}
	\author{Xuejing Zhang,~\IEEEmembership{Student Member,~IEEE,}
	Zishu He,~\IEEEmembership{Member,~IEEE,} and
	Xuepan Zhang
	\thanks{X. Zhang and Z. He are with the University of Electronic Science and Technology of China, Chengdu 611731, China (e-mail: xjzhang7@163.com; zshe@uestc.edu.cn).}
	\thanks{X. P. Zhang is with Qian Xuesen Lab of Space Technology, Beijing 100094, China (e-mail: zhangxuepan@qxslab.cn).}}
\markboth{}
{Shell \MakeLowercase{\textit{et al.}}: Bare Demo of IEEEtran.cls for Journals}
\maketitle

\begin{abstract}
In this paper, the complex-coefficient weight vector orthogonal decomposition
($ \textrm{C}^2\textrm{-WORD} $) algorithm proposed in Part I of this two paper series is
extended to robust sidelobe control and synthesis
with steering vector mismatch.
Assuming that the steering vector uncertainty is norm-bounded,
we obtain the worst-case upper and lower
boundaries of array response.
Then, we devise a robust $ \textrm{C}^2\textrm{-WORD} $ algorithm
to control the
response of a sidelobe point
by precisely adjusting its upper-boundary response level as desired.
To enhance the practicality of the proposed robust 
$ \textrm{C}^2\textrm{-WORD} $
algorithm, we also present detailed analyses on how to determine the
upper norm boundary of steering vector uncertainty under various mismatch
circumstances.
By applying the robust $ \textrm{C}^2\textrm{-WORD} $ algorithm iteratively, a robust sidelobe synthesis approach is developed. 
In this approach, the upper-boundary response is adjusted in a point-by-point manner by successively updating the weight vector.
Contrary to the existing approaches, 
the devised robust $ \textrm{C}^2\textrm{-WORD} $ algorithm has an analytical
expression and can work starting from an arbitrarily-specified
weight vector.
Simulation results are presented to validate the 
effectiveness and good performance of the robust $ \textrm{C}^2\textrm{-WORD} $ 
algorithm.
\end{abstract}


\IEEEpeerreviewmaketitle

\section{Introduction}
\IEEEPARstart{I}{n} the companion paper \cite{robust1},
an array response control scheme named
complex-coefficient weight vector orthogonal decomposition
($ \textrm{C}^2\textrm{-WORD} $) was proposed and analyzed.
In $ \textrm{C}^2\textrm{-WORD} $, the given weight vector
is orthogonally decomposed as two parts.
We realize precise array response control of a given point
by adjusting the combining coefficient (in complex domain) 
of the two orthogonal vectors.
The $ \textrm{C}^2\textrm{-WORD} $ scheme can work from
an arbitrarily-specified weight vector. 
Moreover, it brings
less pattern variations at the uncontrolled points.
As presented in \cite{robust1}, 
one can synthesize desirable patterns by successively
choosing the angle to be controlled and applying $ \textrm{C}^2\textrm{-WORD} $ approach.

In $ \textrm{C}^2\textrm{-WORD} $, the array steering vectors
are assumed to be known exactly.
Under practical circumstances, however, the actual steering
vectors may be different from the assumed (or ideal) ones.
The steering vector uncertainty
can be caused by various factors such as, channel gain-phase mismatch,
element position mismatch and mutual coupling effect.
The existence of steering vector uncertainties
may lead to performance degradation for $ \textrm{C}^2\textrm{-WORD} $
algorithm on array response control and pattern synthesis.

During the past several decades, quite a number of robust algorithms have been developed 
to control array response or synthesize desirable beampatterns
with steering vector uncertainties.
For example, authors of \cite{cwordref2} proposed a powerful robust approach to synthesizing
array patterns with low sidelobes in the presence of unknown
array manifold perturbations. 
This approach optimizes the worst-case performance of sidelobe response
by formulating the robust pattern synthesis problem as a convex programming (CP)
form.
Nevertheless, this method can only synthesize uniform sidelobes
and may not work well if the sidelobe shape is arbitrarily-specified.
A novel robust beampattern synthesis method is proposed in \cite{cwordref3},
where the mutual coupling effect is considered and two
optimization methods are provided.
Nevertheless, the mutual coupling matrix has to be
pre-calculated in this approach before the synthesis process.
Efficient robust broadband antenna array pattern synthesis techniques
in the presence of array imperfections have been
presented in \cite{cwordref4}, where nine different
optimization criteria are provided with
each one having particular advantages and disadvantages for certain
applications.
In contrast to the above deterministic pattern synthesis
methods, there are also some excellent works considering
robust adaptive beamforming with steering vector uncertainties, see \cite{cwordref5,cwordref6,cwordref7,cwordref8,cwordref9}.
In this case, it is usually required to 
shape satisfactory beampatterns
and reject the undesirable interferences.
In addition, it should be pointed out that our discussions
are different from the pattern tolerance analyses in 
\cite{cwordref10,cwordref12,cwordref13,cwordref14,cwordref15}, 
where the weight vector (but not the steering vector)
suffers from perturbation.

In general, the existing methods cannot flexibly control the array response
starting from an arbitrarily-specified weight vector. 
As a result, the weight vector has to be completely redesigned even if only a slight change of the desired pattern is needed.
This motivates us to develop a new array response control
algorithm with steering vector perturbations.
Toward this end, in this paper we modify the 
$ \textrm{C}^2\textrm{-WORD} $ algorithm in \cite{robust1} and
develop a new scheme named robust $ \textrm{C}^2\textrm{-WORD} $.
Starting from an any given weight vector,
the proposed robust $ \textrm{C}^2\textrm{-WORD} $ algorithm
can control the array response level of a single sidelobe point 
when array
suffers from unknown steering vector mismatches.
More specifically, 
assuming that the steering vector 
perturbation is norm-bounded by a known constant,
we first analyze the worst-case (upper and lower) boundaries of
array response level.
Then, given a sidelobe angle to be controlled, its upper
response level, and an arbitrarily-specified weight vector,
we follow the orthogonal decomposition model 
of $ \textrm{C}^2\textrm{-WORD} $ in \cite{robust1} and propose
to accurately control the 
worst-case upper-boundary response level as desired.
As presented later, the robust $ \textrm{C}^2\textrm{-WORD} $ algorithm
offers an analytical expression of weight vector updating
and results less worst-case perturbation on the array response.
In addition, inheriting the advantages of $ \textrm{C}^2\textrm{-WORD} $,
our robust $ \textrm{C}^2\textrm{-WORD} $ approach
results small pattern variations on the uncontrolled points.
To enhance the practicality of the devised algorithm,
we also present how to determine the norm boundary
of steering vector uncertainty, in the cases that array suffers from
channel gain-phase mismatch, element position mismatch
and mutual coupling effect, respectively.
By applying the robust $ \textrm{C}^2\textrm{-WORD} $ algorithm
successively, we devise an effective robust sidelobe synthesis method.
Simulations show that our algorithms work well under various
circumstances.

This paper is organized as follows.
The proposed robust $ \textrm{C}^2\textrm{-WORD} $ algorithm is presented in Section II.
In Section III, some practical considerations are provided to improve
the practicality of robust $ \textrm{C}^2\textrm{-WORD} $.
The application of robust $ \textrm{C}^2\textrm{-WORD} $ to robust
sidelobe synthesis is discussed in Section IV. 
Representative simulations are carried out in Section V and conclusions are drawn in Section VI.

{\textit{Notations:}} Following the notations in \cite{robust1}, we use bold upper-case and lower-case letters to represent
matrices and vectors, respectively.
In particular, we use $ {\bf I} $ to denote the identity matrix.
$ j\triangleq\sqrt{-1} $.
$ (\cdot)^{\mT} $
and $ (\cdot)^{\mH} $ stand for the transpose and Hermitian
transpose, respectively.
$ |\cdot| $ denotes the absolute value and $ \|\cdot\|_2 $ denotes the $ l_2 $ norm.
We use $ {\bf B}(i,l) $ to
stand for the element at the $ i $th row and $ l $th column of matrix $ {\bf B} $.
$ {\bf P}_{\bf Z} $ and $ {\bf P}^{\bot}_{\bf Z} $ represent
the projection matrices onto $ \mathcal{R}({\bf Z}) $
and $ \mathcal{R}^{\bot}({\bf Z}) $, respectively.
$ {\rm Diag}(\cdot) $ represents the diagonal matrix with the components of the input vector as the diagonal elements.
Finally, $ \lambda_{\rm max}(\cdot) $ returns the largest eigenvalue of the input matrix.

\section{Robust Sidelobe Control via $ \textrm{C}^2\textrm{-WORD} $}

\subsection{Robust Sidelobe Control Formulation}
For the ease of later derivations, we first define the normalized magnitude response as
\begin{align}\label{p2001}
V_a(\theta)={|{\bf w}^{\mH}{\bf a}(\theta)|}/{|{\bf w}^{\mH}{\bf a}(\theta_0)|}
\end{align}
where $ {\bf a}(\theta) $ denotes the nominal steering vector as defined in 
{\color{black}Eqn. (2)} in \cite{robust1}, 
$ \theta_0 $ stands for the 
main beam axis. 
Note that the above $ V_a(\theta) $ is different from the normalized power response $ L(\theta,\theta_0) $
defined in {\color{black}Eqn. (1)} in \cite{robust1}. One can readily find that $ V^2_a(\theta)=L(\theta,\theta_0) $.

Clearly, $ V_a(\theta) $ describes the array magnitude response in the absence of array uncertainties.
In practical, however, the steering vector is usually influenced by antenna array imperfections, such as, gain-phase mismatch, element position
mismatch, mutual coupling effect and so on. In this case, the actual steering vector, denoted by $ {\bf b}(\theta) $, is given by
\begin{align}\label{p2002}
{\bf b}(\theta)={\bf a}(\theta)+\Delta(\theta)
\end{align}
where $ \Delta(\theta) $ is the unknown uncertainty that can be varied with $ \theta $. 
The actual normalized magnitude response, denoted by $ V_b(\theta) $, can be expressed as
\begin{align}\label{p2003}
V_b(\theta)={|{\bf w}^{\mH}{\bf b}(\theta)|}/{|{\bf w}^{\mH}{\bf b}(\theta_0)|}
\end{align}
which is different from $ V_a(\theta) $ under normal circumstances.
Note that in \eqref{p2003}, we keep on using
the output of $ {\theta_0} $ as the normalization factor,
although the actual beam axis may have deviated slightly from
$ \theta_0 $ due to the steering vector uncertainties.
In robust sidelobe control, we consider how to make the
actual magnitude response $ V_b(\theta) $ lower than specific levels
in certain sidelobe regions.

\subsection{Boundary Analysis on Array Response}
To proceed, we first present a boundary analysis on the magnitude response $ V_b(\theta) $.
To do so, we reasonably suppose that the uncertainty $ \Delta(\theta) $ is norm-bounded as
\begin{align}\label{p2005}
\|\Delta(\theta)\|_2\leq\varepsilon(\theta)
\end{align}
where $ \varepsilon(\theta) $ is a known constant at $ \theta $.
Then, according to the triangle inequality property, one gets
\begin{align}\label{p2006}
|{\bf w}^{\mH}{\bf b}(\theta)|&=|{\bf w}^{\mH}({\bf a}(\theta)+\Delta(\theta))|\nonumber\\
&\leq|{\bf w}^{\mH}{\bf a}(\theta)|+|{\bf w}^{\mH}\Delta(\theta)|\nonumber\\
&\leq|{\bf w}^{\mH}{\bf a}(\theta)|+\|{\bf w}\|_2{\cdot}\|\Delta(\theta)\|_2\nonumber\\
&\leq|{\bf w}^{\mH}{\bf a}(\theta)|+\varepsilon(\theta)\|{\bf w}\|_2.
\end{align}
Similarly,
\begin{align}\label{p2007}
|{\bf w}^{\mH}{\bf b}(\theta)|&=|{\bf w}^{\mH}({\bf a}(\theta)+\Delta(\theta))|\nonumber\\
&\geq|{\bf w}^{\mH}{\bf a}(\theta)|-|{\bf w}^{\mH}\Delta(\theta)|\nonumber\\
&\geq|{\bf w}^{\mH}{\bf a}(\theta)|-\|{\bf w}\|_2{\cdot}\|\Delta(\theta)\|_2\nonumber\\
&\geq|{\bf w}^{\mH}{\bf a}(\theta)|-\varepsilon(\theta)\|{\bf w}\|_2.
\end{align}
Combining \eqref{p2006} and \eqref{p2007}, one can readily find that
\begin{align}\label{p2008}
V_b(\theta)&=\dfrac{|{\bf w}^{\mH}{\bf b}(\theta)|}{|{\bf w}^{\mH}{\bf b}(\theta_0)|}\nonumber\\
&=\dfrac{|{\bf w}^{\mH}({\bf a}(\theta)+\Delta(\theta))|}
{|{\bf w}^{\mH}({\bf a}(\theta_0)+\Delta(\theta_0))|}\nonumber\\
&\leq\dfrac{|{\bf w}^{\mH}{\bf a}(\theta)|+\varepsilon(\theta)\|{\bf w}\|_2}
{|{\bf w}^{\mH}{\bf a}(\theta_0)|-\varepsilon(\theta_0)\|{\bf w}\|_2}\nonumber\\
&=\dfrac{V_a(\theta)+\varepsilon(\theta)\cdot\|{\bf w}\|_2/|{\bf w}^{\mH}{\bf a}(\theta_0)|}
{1-\varepsilon(\theta_0)\cdot\|{\bf w}\|_2/|{\bf w}^{\mH}{\bf a}(\theta_0)|}\nonumber\\
&\triangleq V_{u}(\theta)
\end{align}
and
\begin{align}\label{p2120}
V_b(\theta)&\geq\dfrac{|{\bf w}^{\mH}{\bf a}(\theta)|-\varepsilon(\theta)\|{\bf w}\|_2}
{|{\bf w}^{\mH}{\bf a}(\theta_0)|+\varepsilon(\theta_0)\|{\bf w}\|_2}\nonumber\\
&=\dfrac{V_a(\theta)-\varepsilon(\theta)\cdot\|{\bf w}\|_2/|{\bf w}^{\mH}{\bf a}(\theta_0)|}
{1+\varepsilon(\theta_0)\cdot\|{\bf w}\|_2/|{\bf w}^{\mH}{\bf a}(\theta_0)|}\nonumber\\
&\triangleq V_{l}(\theta).
\end{align}
Compactly, we have
\begin{align}\label{key100}
0\leq V_{l}(\theta)\leq V_{b}(\theta)\leq V_{u}(\theta)
\end{align}
where $ V_{u}(\theta) $ and $ V_{l}(\theta) $ stand for the
worst-case upper and lower boundaries of magnitude response, respectively. 
According to \eqref{key100}, the actual response $ V_b(\theta) $
fluctuates in the range $ [V_l(\theta),V_u(\theta)] $.
In addition, it should be noted that we have implicitly assumed in \eqref{p2008} that
\begin{align}\label{p2110}
{|{\bf w}^{\mH}{\bf a}(\theta_0)|-\varepsilon(\theta_0)\|{\bf w}\|_2}>0.
\end{align}
Otherwise, it leads to $ V_{u}(\theta)<0 $ and \eqref{p2008} does not hold true.

\subsection{Robust One-Point Sidelobe Control Formulation}
In the preceding subsection, a boundary analysis on the array response is presented. In this subsection, 
we formulate the problem of robust one-point
sidelobe control, i.e., 
making the response level of a
given sidelobe point lower than specific value in the presence of steering
vector uncertainties.

More specifically, 
denote by 
$ V_d(\theta) $ the desired magnitude upper beampattern.
Give a previous weight vector $ {\bf w}_{k-1} $ and
a sidelobe angle $ \theta_k $ to be controlled.
It is required to find a new weight vector $ {\bf w}_k $ that makes the
actual (magnitude) response level of $ \theta_k $ 
lower than $ {V_d}(\theta_k) $.
To simplify notations, in sequel we follow the usages of
$ V_a(\theta) $, $ V_b(\theta) $, $ V_u(\theta) $ and
$ V_l(\theta) $ defined in the two preceding subsections, and designate them to
stand for the counterparts of $ {\bf w}_k $.
Then,
the problem of
one-point robust sidelobe control can be formulated as
\begin{subequations}\label{p2004}
	\begin{align}
	{\rm find}&~~~{\bf w}_k\\
	\label{p2219}{\rm subject~to}&~~~V_{b}(\theta_k)\leq{V_d}(\theta_k).
	\end{align}
\end{subequations}
Note that there exists unknown perturbations on the steering vector, see \eqref{p2002}. As a result, it may not be easy to adjust 
$ V_{b}(\theta_k) $ as desired.

To tackle problem \eqref{p2004}, we
recall \eqref{key100} and formulate a conservative version
of \eqref{p2004} as
\begin{subequations}\label{p2009}
	\begin{align}
	{\rm find}&~~~{\bf w}_k\\
	\label{p2109}{\rm subject~to}&~~~V_{u}(\theta_k)\leq{V_d}(\theta_k)
	\end{align}
\end{subequations}
where the maximum possible response level at $ \theta_k $ 
(i.e., $ V_{u}(\theta_k) $) is restricted to be lower than $ {V_d}(\theta_k) $. 
Since $ V_{b}(\theta_k) $ is not greater than $ V_{u}(\theta_k) $,
one learns that the original constraint \eqref{p2219} is satisfied
if only \eqref{p2109} holds true.

One possible way to make constraint \eqref{p2109} qualified is to take
$ V_{u}(\theta_k) $ as its minimum, which may be close to zero. 
By doing so, 
the actual response $ V_{b}(\theta_k) $ would also approach to zero
because of the constraint \eqref{key100}.
As a result, it may broaden the mainlobe of 
$ V_{b}(\theta) $ and/or lower the resulting WNG.
To alleviate this drawback, a high value of $ V_b(\theta_k) $
is expected under the condition that \eqref{p2109} is satisfied.
As aforementioned, $ V_b(\theta_k) $ fluctuates in
the range $ [V_l(\theta_k),V_u(\theta_k)] $.
Then, a reasonable way to elevate $ V_b(\theta_k) $
is to lift both the lower-boundary response level $ V_l(\theta_k) $
and the upper-boundary response level $ V_u(\theta_k) $.
According to \eqref{p2109},
the maximum of $ V_u(\theta_k) $ is $ V_d(\theta_k) $,
then we can improve the general level of $ V_b(\theta_k) $ by 
fixing $ V_u(\theta_k) $ as $ V_d(\theta_k) $
and then solving the following optimization problem:
\begin{subequations}\label{p21236}
	\begin{align}
	\max_{{\bf w}_k}&~~~V_{l}(\theta_k)\\
	{\rm subject~to}&~~~V_{u}(\theta_k)={V_d}(\theta_k).
	\end{align}
\end{subequations}

For the given $ V_a(\theta_k) $, $ \varepsilon(\theta_0) $ and 
$ \varepsilon(\theta_k) $, it is not hard to observe from \eqref{p2008} and \eqref{p2120} that
\begin{align}
V_u(\theta_k)-V_l(\theta_k) \propto 
\dfrac{\|{\bf w}_k\|_2}{|{\bf w}^{\mH}_k{\bf a}(\theta_0)|}.
\end{align}
Thus, one can reformulate problem \eqref{p21236} as:
\begin{subequations}\label{p2126}
	\begin{align}
	\max_{{\bf w}_k}&~~~G({\bf w}_k)\\
	{\rm subject~to}&~~~V_{u}(\theta_k)={V_d}(\theta_k)
	\end{align}
\end{subequations}
where
\begin{align}
G({\bf w})\triangleq\dfrac{|{\bf w}^{\mH}{\bf a}(\theta_0)|^2}
{\|{\bf w}\|^2_2}
\end{align}
stands for the white noise gain (WNG) in the
absence of steering vector uncertainties, and has also been defined in {\color{black}Eqn. (20)} in \cite{robust1}.
As a matter of fact, since $ \|{\bf w}_k\|_2/|{\bf w}^{\mH}_k{\bf a}(\theta_0)| $ is
directly proportional to the
pattern perturbations, i.e., $ V_u(\theta_k)-V_a(\theta_k) $ and
$ V_a(\theta_k)-V_l(\theta_k) $,
the solution of problem \eqref{p2126}
obtains small pattern perturbations for the given
$ \varepsilon(\theta_0) $ and 
$ \varepsilon(\theta_k) $.

Recalling the definition of $ V_u(\theta_k) $ in \eqref{p2008}, we can
reformulate the robust one-point sidelobe control
problem \eqref{p2126} as
\begin{subequations}\label{p2127}
	\begin{align}
	\max_{{\bf w}_k}&~~~G({\bf w}_k)\\
	\label{p2118}{\rm subject~to}&~~~V_a(\theta_k)=V_{\bf w}(\theta_k)
	\end{align}
\end{subequations}
where $ V_{\bf w}(\theta) $ is defined as
\begin{align}\label{p2011}
V_{\bf w}(\theta)\triangleq{V_d}(\theta)-{\gamma(\theta)\|{\bf w}_k\|_2}/
{|{\bf w}^{\mH}_k{\bf a}(\theta_0)|}
\end{align}
with
$ \gamma(\theta)\triangleq{V_d}(\theta)\varepsilon(\theta_0)+\varepsilon(\theta) $.
Clearly, the non-convex problem \eqref{p2127} maximizes WNG
with specific constraint on the (ideal) response level of $ \theta_k $.
This is similar to the response control problem formulated in
Part I \cite{robust1}.
Nevertheless, 
different from the array response control problem
discussed in \cite{robust1},
it should be noted that the right side of
constraint \eqref{p2118}, i.e., $ V_{\bf w}(\theta_k) $, depends on the
optimization variable $ {\bf w}_k $ as well.
In addition, the previous weight vector $ {\bf w}_{k-1} $ is
not taken into consideration in the formulating problem \eqref{p2127}.
As a result, it may lead to large pattern variations at the uncontrolled points,
comparing to the previous beampattern response.
Recalling that the $ \textrm{C}^2 $-$ \textrm{WORD} $ scheme devised in Part I \cite{robust1} maximizes WNG and {\color{black}results small pattern variations},
this provides some inspiration to use $ \textrm{C}^2 $-$ \textrm{WORD} $ algorithm in \cite{robust1} to realize one-point sidelobe control, as detailed in the
next subsection.

\subsection{Robust $ \textrm{C}^2 $-$ \textrm{WORD} $ Algorithm}
In this subsection, we propose a new method to
realize robust one-point sidelobe control and name
it as
robust $ \textrm{C}^2 $-$ \textrm{WORD} $ algorithm.
The devised algorithm is built on the foundation
of $ \textrm{C}^2 $-$ \textrm{WORD} $ scheme developed in \cite{robust1} and can result small pattern variations
at the uncontrolled points.

To begin with, we recall $ \textrm{C}^2 $-$ \textrm{WORD} $ scheme in \cite{robust1},
and incorporate a new constraint into \eqref{p2127} as
\begin{subequations}\label{p21127}
	\begin{align}
	\max_{{\beta}_k}&~~~G({\bf w}_k)\\
	\label{p21118}{\rm subject~to}&~~~V_a(\theta_k)=V_{\bf w}(\theta_k)\\
\label{p211118}&~~~{\bf w}_{k}=
	\begin{bmatrix}{\bf{w}}_{\bot}&{\bf{w}}_{\Arrowvert}\end{bmatrix} 
	\begin{bmatrix}1& {\beta}_{k}\end{bmatrix}^{\mT},~{\beta}_{k}\in\mathbb{C}
	\end{align}
\end{subequations}
where the constraint of orthogonal decomposition has been added
in \eqref{p211118},
with $ {\bf{w}}_{\bot} $ and $ {\bf{w}}_{\Arrowvert} $ satisfying
\begin{align}\label{word01}
{\bf{w}}_{\bot}\triangleq{\bf{P}}^{\bot}_{[{\bf{a}}(\theta_{k})]}{\bf{w}}_{k-1},~~
{\bf{w}}_{\Arrowvert}\triangleq{\bf{P}}_{[{\bf{a}}(\theta_{k})]}{\bf{w}}_{k-1}.
\end{align}
Once the optimal $ {\beta}_{k,\star} $ of \eqref{p21127} is obtained,
we can express the ultimate weight vector $ {\bf w}_{k} $ as
\begin{align}\label{p21015}
{\bf w}_{k}=
\begin{bmatrix}{\bf{w}}_{\bot}&{\bf{w}}_{\Arrowvert}\end{bmatrix} 
\begin{bmatrix}1& {\beta}_{k,\star}\end{bmatrix}^{\mT}.
\end{align}
It should be emphasized that the resulting weight vector \eqref{p21015}
may not be the  global optimal solution of problem \eqref{p2127},
since we have assigned a new constraint in problem \eqref{p21127}.
In spite of that, we will show later that the obtained $ {\bf w}_k $
in \eqref{p21015} performs well on robust sidelobe control
with small pattern variations at the uncontrolled points.
The remaining problem is how to determine the optimal
$ {\beta}_{k,\star} $ of \eqref{p21127}, as presented next.


%
%

The $ \textrm{C}^2 $-$ \textrm{WORD} $ scheme in \cite{robust1}
is able to precisely control the array power response of a given point.
For the sake of subsequent convenience,
we first define 
\begin{align}\label{rhoa}
{\rho}_a\triangleq V^2_a(\theta_k)
\end{align}
which represents the ideal power response level at $ \theta_k $.
Then, for the given $ \theta_0 $, $ \theta_k $, $ {\bf{w}}_{k-1} $, 
$ {V_d}(\theta_k) $, $ \varepsilon(\theta_0) $ and $ \varepsilon(\theta_k) $,
we can indirectly determine the optimal $ \beta_{k,\star} $ by
finding the corresponding $ {\rho}_a $ in problem \eqref{p21127}.

After some calculation, one can see that the corresponding $ {\rho}_a $ of problem \eqref{p21127}
satisfies the following quartic polynomial:
\begin{align}\label{qus}
A^2{\rho}^4_{a}&+(2AC-B^2){\rho}^3_{a}+(2AE-2BD+C^2){\rho}^2_{a}\nonumber\\
&~~~~~~~~~+(2CE-D^2){\rho}_{a}+E^2=0.
\end{align}
The derivation of \eqref{qus} is presented
in Appendix A, where the expressions of $ A,B,C,D,E $ are also specified, see \eqref{ABCDE}.
In fact, there are four candidates of $ {\rho}_a $ satisfying \eqref{qus},
and they can be analytically expressed, see the following Lemma to find
their specific expressions.

%
%
%
%

\newtheorem{theoreml}{Lemma}
\begin{theoreml}
	The four roots $ x_i $ ($ i=1,2,3,4 $) for the following general quartic equation
	\begin{align}
	ax^4+bx^3+cx^2+dx+e=0,~(a\neq 0)
	\end{align}
	are given by 
	\begin{subequations}\label{solutionx}
		\begin{align}
		x_1&=-\dfrac{b}{4a}+S+\dfrac{\sqrt{-4S^2-2p-\frac{q}{S}}}{2}\\
		x_2&=-\dfrac{b}{4a}+S-\dfrac{\sqrt{-4S^2-2p-\frac{q}{S}}}{2}\\
		x_3&=-\dfrac{b}{4a}-S+\dfrac{\sqrt{-4S^2-2p+\frac{q}{S}}}{2}\\
		x_4&=-\dfrac{b}{4a}-S-\dfrac{\sqrt{-4S^2-2p+\frac{q}{S}}}{2}
		\end{align}
	\end{subequations}
	where
	\begin{subequations}
		\begin{align}
		p&\triangleq\dfrac{8ac-3b^2}{8a^2}\\
		q&\triangleq\dfrac{b^3-4abc+8a^2d}{8a^3}\\
		S&\triangleq
		\dfrac{\sqrt{-\frac{2}{3}p+\frac{1}{3a}\left(Q+\frac{\zeta_0}{Q}\right)}}{2}
		\end{align}
	\end{subequations}		
	with 
	\begin{subequations}
		\begin{align}
		Q&\triangleq
		\sqrt[\uproot{6}3]{\left({\zeta_1+
				\sqrt{\zeta^2_1-4\zeta^3_0}}\right)\big/{2}}\\
		\zeta_0&\triangleq c^2-3bd+12ae\\
			\zeta_1&\triangleq 2c^3-9bcd+27b^2e+27ad^2-72ace.
		\end{align}
	\end{subequations}
\end{theoreml}
\begin{IEEEproof}
	See \cite{roots}.
\end{IEEEproof}

According to \eqref{solutionx} in Lemma 1, we can similarly obtain the four solutions of quartic equation \eqref{qus}, and denote them as 
$ \rho_{a,1},\rho_{a,2},\rho_{a,3},\rho_{a,4} $, respectively.
Recalling our previous discussions, we
note that the qualified $ \rho_a $ is real-valued and satisfies
\begin{align}
{\rho}_{a}\in[0,{V^2_d}(\theta_k)).
\end{align}
Then, to obtain the maximal $ V_a(\theta_k) $, we can obtain the final 
$ \rho_{a,\star} $ by solving the following simple problem:
\begin{subequations}\label{p221127}
	\begin{align}
	\max_{\rho_a}&~~~\rho_a\\
	\label{p213118}{\rm subject~to}&~~~\rho_a\in\{\rho_{a,1},\rho_{a,2},\rho_{a,3},\rho_{a,4}\}\\
	\label{p2131118}&~~~{\rho}_{a}\in[0,{V^2_d}(\theta_k)).
	\end{align}
\end{subequations}
Once the optimal $ \rho_{a,\star} $ is determined, we can remove
the constraint \eqref{p21118} and reformulate problem
\eqref{p21127} as
\begin{subequations}\label{p321127}
	\begin{align}
	\max_{{\beta}_k}&~~~G({\bf w}_k)\\
	\label{p2123118}{\rm subject~to}&~~~L_k(\theta_k,\theta_0)=\rho_{a,\star}\\
	\label{p21221118}&~~~{\bf w}_{k}=
	\begin{bmatrix}{\bf{w}}_{\bot}&{\bf{w}}_{\Arrowvert}\end{bmatrix} 
	\begin{bmatrix}1& {\beta}_{k}\end{bmatrix}^{\mT},~{\beta}_{k}\in\mathbb{C}
	\end{align}
\end{subequations}
where $ L_{k}(\theta,\theta_0)=
{| {\bf{w}}^{\mH}_{k}{\bf{a}}(\theta) |^2}/
{| {\bf{w}}^{\mH}_{k}{\bf{a}}(\theta_0) |^2} $. 
The above problem \eqref{p321127} has an analytical solution,
see Proposition 2 in Part I \cite{robust1} for details.
Thus, we can
obtain the ultimate $ \beta_{k,\star} $ and its corresponding
weight vector $ {\bf w}_k $ in \eqref{p21015}.
This completes the robust one-point response control at a 
given sidelobe point. Finally, we summarize
the proposed robust $ \textrm{C}^2 $-$ \textrm{WORD} $ algorithm
in Algorithm \ref{fact33orps}.

\begin{algorithm}[!t]
	\caption{Robust $ \textrm{C}^2\textrm{-WORD} $ Algorithm}\label{fact33orps}
	\begin{algorithmic}[1]
	\State prescribe beam axis $ \theta_0 $ and index $ k $, give the previous weight vector $ {\bf w}_{k-1} $, 
	sidelobe angle $ {\theta_{k}} $, the desired (magnitude) upper response level $ {V}_d(\theta_k) $ and the steering vector uncertainty boundaries $ \varepsilon(\theta_0) $ and $ \varepsilon(\theta_k) $
	\State construct quartic equation \eqref{qus} and find its solutions
	(i.e.,
	$ {\rho}_{a,1} $, $ {\rho}_{a,2} $, $ {\rho}_{a,3} $, $ {\rho}_{a,4} $) from Lemma 1
	\State determine the ultimate $ \rho_{a,\star} $ by solving problem \eqref{p221127}
	\State apply $ \textrm{C}^2\textrm{-WORD} $ algorithm to adjust the ideal
	array response of $ \theta_k $ to $ {\rho}_{a,\star} $ and obtain the corresponding
	$ {\beta}_{k,\star} $, see Algorithm 1 in \cite{robust1}
	\State output the new weight vector $ {\bf w}_k $ in \eqref{p21015}		
\end{algorithmic}
\end{algorithm}

\subsection{Restriction Between $ {V_d}(\theta_k) $ and $ \varepsilon(\theta_k) $}
In the preceding subsection, we use the $ \textrm{C}^2 $-$ \textrm{WORD} $
scheme to realize robust sidelobe control at a pre-assigned angle $ \theta_k $.
It should be pointed out that there exists an implicit
restriction between the minimum reachable upper response level $ {V_d}(\theta_k) $ and the
steering vector uncertainty norm boundary $ \varepsilon(\theta_k) $, as investigated next.

To begin with, we note that $ V_a(\theta_k)\geq 0 $.
Since $ V_a(\theta_k) $ is generally in proportional to $ V_u(\theta_k) $,
we can obtain the minimum reachable level of $ {V_d}(\theta_k) $ (denoted as
$ \overline{V}_d(\theta_k) $) by setting $ V_a(\theta_k)=0 $. 
Recalling \eqref{p2011} and \eqref{p21118},
$ \overline{V}_d(\theta_k) $ is given by
\begin{align}\label{minu}
\overline{V}_d(\theta_k)&=\gamma(\theta_k)\|{\bf w}_k\|_2/
{|{\bf w}^{\mH}_k{\bf a}(\theta_0)|}\nonumber\\
&=(\overline{V}_d(\theta_k)\varepsilon(\theta_0)+\varepsilon(\theta_k))\cdot\|{\bf w}_k\|_2/
{|{\bf w}^{\mH}_k{\bf a}(\theta_0)|}\nonumber\\
&=(\overline{V}_d(\theta_k)\varepsilon(\theta_0)+\varepsilon(\theta_k))\cdot\|{\bf w}_{\bot}\|_2/
{|{\bf w}^{\mH}_{\bot}{\bf a}(\theta_0)|}
\end{align}
where we have utilized the fact that
$ \beta_{\star}=0 $ and $ {\bf w}_{k}={\bf{w}}_{\bot} $ when
$ V_a(\theta_k)=0 $ applies.
According to \eqref{minu}, $ {V_d}(\theta_k) $ should be taken
to satisfy:
\begin{align}\label{key0123}
{V_d}(\theta_k)\geq\overline{V}_d(\theta_k)=\dfrac{\varepsilon(\theta_k)\|{\bf w}_{\bot}\|_2}
{|{\bf w}^{\mH}_{\bot}{\bf a}(\theta_0)|-\varepsilon(\theta_0)\|{\bf w}_{\bot}\|_2}
\end{align}
which specifies the restriction between $ {V_d}(\theta_k) $ and $ \varepsilon(\theta_k) $
for the given $ \varepsilon(\theta_0) $ and $ {\bf a}(\theta_0) $.
Clearly, the less $ \varepsilon(\theta_k) $ is,
the lower level $ {V_d}(\theta_k) $ can be taken.

Note from \eqref{key0123} that the minimum reachable level
of $ V_d(\theta_k) $ (i.e, $ \overline{V}_d(\theta_k) $)
depends on the previous weight vector 
$ {\bf w}_{k-1} $ as well.
Utilizing the Cauchy-Schwarz inequality that 
$ \|{\bf w}_{\bot}\|_2\|{\bf a}(\theta_0)\|_2\geq|{\bf w}^{\mH}_{\bot}{\bf a}(\theta_0)| $, one can further obtain:
\begin{align}\label{chi}
{V}_d(\theta_k)\geq\overline{V}_d(\theta_k)\geq\dfrac{\varepsilon(\theta_k)}
{\|{\bf a}(\theta_0)\|_2-\varepsilon(\theta_0)}\triangleq\chi(\theta_k)
\end{align}
where the introduced $ \chi(\theta_k) $ is
independent of the weight vector
and specifies a lower boundary
of $ \overline{V}_d(\theta_k) $.
Note that the resulting $ \chi(\theta_k) $ in \eqref{chi} gives
a general value of the minimum achievable level
of $ V_d(\theta_k) $,
although it may not be 
reached for an arbitrarily-specified previous weight $ {\bf w}_{k-1} $.

\section{Practical Consideration}
In the preceding section, a robust $ \textrm{C}^2 $-$ \textrm{WORD} $
algorithm is devised to adjust
the response level of a pre-assigned sidelobe point with
steering vector perturbation.
As aforementioned, the steering vector uncertainty $ \Delta(\theta) $ 
is assumed to be norm-bounded by a known constant $ \varepsilon(\theta) $.
To enhance the practicality of robust $ \textrm{C}^2 $-$ \textrm{WORD} $
algorithm, we next consider how to determine $ \Delta(\theta) $ and the corresponding
$ \varepsilon(\theta) $ in practical applications with some reasonable assumptions.

In this paper, we assume that the array suffers from
channel gain-phase mismatch, element position mismatch, 
mutual coupling effect or their superpositions.
On this basis, we can model the actual steering vector $ {\bf b}(\theta) $
in \eqref{p2002} as
\begin{align}\label{CC}
{\bf b}(\theta)={\bf C}(\theta){\bf a}(\theta)=
{\bf a}(\theta)+
\underbrace{[{\bf C}(\theta)-{\bf I}]{\bf a}(\theta)}_{\Delta(\theta)}
\end{align}
where $ {\bf C}(\theta) $ is a certain matrix whose elements may vary with $ \theta $.
Let us define
\begin{align}\label{CE}
{{\bf E}(\theta)}\triangleq{{\bf C}(\theta)-{\bf I}}.
\end{align}
Then according to \eqref{CC}, we have
$ \Delta(\theta)={{\bf E}(\theta)}{\bf a}(\theta) $,
from which one can further derive that
\begin{align}\label{q3001}
\|\Delta(\theta)\|_2=\|{\bf E}(\theta){\bf a}(\theta)\|_2
\leq\|{\bf E}(\theta)\|_2\|{\bf a}(\theta)\|_2
\end{align}
where $ \|{\bf E}(\theta)\|_2 $ is the spectral matrix norm \cite{refmat} of $ {\bf E}(\theta) $ 
satisfying
\begin{align}\label{spectralnorm}
\|{\bf E}(\theta)\|_2=\sqrt{\lambda_{\rm max}({\bf E}^{\mH}(\theta){\bf E}(\theta))}.
\end{align}
Clearly, if $ {\bf E}(\theta) $ is known or can be estimated,
we can set the corresponding $ \varepsilon(\theta) $ in \eqref{p2005}
as
\begin{align}\label{varesp2lon}
\varepsilon(\theta)=\|{\bf E}(\theta)\|_2\|{\bf a}(\theta)\|_2.
\end{align}
However, this is not a common occurrence, since
$ {\bf E}(\theta) $ (or $ {\bf C}(\theta) $) is usually
a random matrix with certain statistical
model for its entries \cite{effectnorm}. In this case, we can determine $ \varepsilon(\theta) $ by
\begin{align}\label{varesplon}
\varepsilon(\theta)=\|{\bf a}(\theta)\|_2\cdot\max~\|{\bf E}(\theta)\|_2.
\end{align}

We next present some specific scenarios, in which
the steering vector uncertainty boundary $ \varepsilon(\theta) $ can
be analytically expressed according to \eqref{varesplon}.
For the sake of simplicity, we only consider linear arrays,
although the extension to more complicated configurations
are straightforward.

%

\subsection{Channel Gain-phase Mismatch}
To begin with, we consider the channel gain-phase mismatch \cite{phaseerror1,phaseerror2,phaseerror3}.
In this case, we have
\begin{align}
{\bf C}(\theta)={\rm Diag}([1,g_2e^{j\varphi_2},\cdots,g_Ne^{j\varphi_N}])
\end{align}
where $ g_n $ and $ \varphi_n $ stand for the channel gain and phase errors of the
$ n $th element, respectively,
$ n=2,\cdots,N $. 
Note that the measurements have been normalized
by that of the first element. Accordingly, $ {\bf E}(\theta) $ can be expressed as
\begin{align}
{\bf E}(\theta)={\rm Diag}([0,g_2e^{j\varphi_2}-1,\cdots,g_Ne^{j\varphi_N}-1]).
\end{align}
Recalling \eqref{spectralnorm}, we have
\begin{align}\label{key125}
\max~\|{\bf E}(\theta)\|_2=\max_{n=2,\cdots,N}~|g_ne^{j\varphi_n}-1|.
\end{align}
Suppose that $ g_n $ and $ \varphi_n $ are randomly distributed
in certain regions as
\begin{align}
g_n\in[g_{n,l},g_{n,u}],~~\varphi_n\in[\varphi_{n,l},\varphi_{n,u}],~n=2,\cdots,N
\end{align}
where $ g_l $, $ g_u $, $ \varphi_l $ and $ \varphi_u $ are 
corresponding boundaries.
On this basis, we can obtain from \eqref{key125} that
\begin{align}
\max~\|{\bf E}(\theta)\|_2=\!\!\!\!\!\!\!\!\max_{n\in\{2,\cdots,N\},\tau\in\{l,u\},\varsigma\in\{l,u\}}\!\!|g_{n,\tau}e^{j\varphi_{n,\varsigma}}-1|
\triangleq\delta_{1}.\nonumber
\end{align}
According to \eqref{varesplon}, one can set $ \varepsilon(\theta) $ as
\begin{align}\label{eqn0998}
\varepsilon(\theta)=\|{\bf a}(\theta)\|_2\cdot\delta_{1}.
\end{align}

\subsection{Element Position Mismatch}
We next analyze the
steering vector uncertainty $ \Delta(\theta) $ and determine its corresponding 
$ \varepsilon(\theta) $ 
when array elements suffer from position uncertainties \cite{position1}.
In this case, $ {\bf C}(\theta) $ is given by
\begin{align}
{\bf C}(\theta)={\rm Diag}([1,
e^{j2\pi\alpha_{2}{\rm sin}(\theta)/\lambda},\cdots,
e^{j2\pi{\alpha}_{N}{\rm sin}(\theta)/\lambda}])
\end{align}
where $ \lambda $ stands for wavelength,
$ \alpha_n $ represents the position deviation of the $ n $th
element from its ideal location $ d_n $, $ n=2,\cdots,N $. 
According to \eqref{CE}, one can express $ {\bf E}(\theta) $ as
\begin{align}
{\bf E}(\theta)={\rm Diag}([0,e^{j2\pi{\alpha}_{2}{\rm sin}(\theta)/\lambda}-1,
\cdots,e^{j2\pi{\alpha}_{N}{\rm sin}(\theta)/\lambda}-1]).\nonumber
\end{align}
Suppose that $ \alpha_n $ is randomly distributed in the range as
\begin{align}
\alpha_n\in[\alpha_{n,l},\alpha_{n,u}],~n=2,\cdots,N
\end{align}
where $ \alpha_{n,l} $ and $ \alpha_{n,u} $ are known boundaries.
Then, we can obtain the following result about $ {\bf E}(\theta) $, i.e.,
\begin{align}
\max~\|{\bf E}(\theta)\|_2=\!\!\!\!\!\!\!\!\max_{n\in\{2,\cdots,N\},\tau\in\{l,u\}}\!\!
|e^{j2\pi{\alpha}_{n,\tau}{\rm sin}(\theta)/\lambda}-1|
\triangleq\delta_{2}(\theta).\nonumber
\end{align}
Different from the $ \delta_{1} $ in Section III.A,
the above $ \delta_{2}(\theta) $ is directionally dependent.
Finally, according to \eqref{varesplon}, we can set 
$ \varepsilon(\theta) $ as
\begin{align}\label{epsi02}
\varepsilon(\theta)=\|{\bf a}(\theta)\|_2\cdot\delta_{2}(\theta).
\end{align}

\subsection{Mutual Coupling Effect}
Now we consider the steering vector uncertainty
arising from mutual coupling effect \cite{mutual1,mutual2,mutual3}.
Following \cite{effectnorm}, we only consider the electromagnetic
coupling between nonadjacent elements of a linear array, and express
the mutual coupling matrix $ {\bf C}(\theta) $ as
\begin{align}
{\bf C}(\theta)=
\begin{bmatrix}
1 		& \xi\cdot z_{1,2} & 0 		& \cdots & 0\\
\xi\cdot z_{2,1} & 1		  & \xi\cdot z_{2,3} & \ddots &\vdots\\
0 		& \xi\cdot z_{3,2} & 1 		& \ddots &0\\
\vdots  & \ddots  & \ddots  & 1      &\xi\cdot z_{N-1,N}\\
0 		& \cdots & 0 		& \xi\cdot z_{N,N-1} &1\\
\end{bmatrix}\nonumber
\end{align}
which is complex symmetry.
In $ {\bf C}(\theta) $, $ \xi $ is the known
isolation between channels,
$ z_{i,j} $ are random variables with a fixed magnitude $ |z_{i,j}|=1 $.
Recalling \eqref{CE}, we have
\begin{align}
{\bf E}(\theta)=\xi\cdot
\begin{bmatrix}
0		&  z_{1,2} & 0 		& \cdots & 0\\
z_{2,1} & 0		  & z_{2,3} & \ddots &\vdots\\
0 		& z_{3,2} & 0 		& \ddots &0\\
\vdots  & \ddots  & \ddots  & 0      & z_{N-1,N}\\
0 		& \cdots & 0 		& z_{N,N-1} &0\\
\end{bmatrix}.
\end{align}
According to the Gershgorin circle theorem \cite{gct}, i.e.,
\begin{align}
|\lambda_{\rm max}({\bf D})|\leq\max_{p=1,\cdots,P}~\sum_{l=1}^{L}|{\bf D}({p,l})|
\end{align}
where $ {\bf D} $ is a $ P\times L $ matrix, it can be concluded that
\begin{align}
\max~\|{\bf E}(\theta)\|_2=\sqrt{\lambda_{\rm max}({\bf E}^{\mH}(\theta){\bf E}(\theta))}
\leq 2\xi\triangleq\delta_{3}.
\end{align}
From \eqref{varesplon}, we can set $ \varepsilon(\theta) $ as
\begin{align}\label{upper3}
\varepsilon(\theta)=\|{\bf a}(\theta)\|_2\cdot\delta_{3}=2\xi\|{\bf a}(\theta)\|_2.
\end{align}

\section{Robust Sidelobe Synthesis}
In this section, 
we introduce
the application of robust $ \textrm{C}^2\textrm{-WORD} $ algorithm
to sidelobe synthesis with steering vector imperfections.
The general strategy is similar to the concept of pattern synthesis
using $ \textrm{C}^2\textrm{-WORD} $ in Part I \cite{robust1}.
However, different from the pattern synthesis approach in \cite{robust1},
we realize robust sidelobe synthesis by
successively adjusting the worst-case upper-boundary
magnitude pattern (i.e., $ V_u(\theta) $), but not the ideal
beampattern $ V_a(\theta) $.
For the sake of clarity, in sequel we incorporate 
the subscript $ k $ into $ V_a(\theta) $ and $ V_u(\theta) $,
and use $ V_{a,k}(\theta) $ and $ V_{u,k}(\theta) $ to stand for
the counterparts of $ {\bf w}_k $.
%
%

More precisely, an initial ideal pattern $ V_{a,0}(\theta) $ and the corresponding
worst-case upper-boundary pattern $ V_{u,0}(\theta) $ are 
firstly obtained from \eqref{p2001} and \eqref{p2008}, respectively, by setting the initial weight vector
as $ {\bf w}_{0} $. Then, following the angle selection strategy in Part I \cite{robust1},
we choose an angle $ \theta_1 $, at which 
$ V_{u,0}(\theta) $ has peak point and produces a higher level than the desired upper pattern
$ V_d(\theta) $.
Next, the robust $ \textrm{C}^2\textrm{-WORD} $ scheme is applied to modify the weight vector $ {\bf w}_{0} $ to $ {\bf w}_{1} $, by setting the output of $ V_{u,1}(\theta_1) $ as $ V_d(\theta_1) $.
Similarly, by comparing $ V_{u,1}(\theta) $ with $ V_d(\theta) $, a second angle $ \theta_2 $, at which the response is needed to be adjusted, is selected. An updated weight vector $ {\bf w}_{2} $ can thus be achieved via robust $ \textrm{C}^2\textrm{-WORD} $. The above procedure is carried out successively 
once the sidelobe responses of $ V_{u,k}(\theta) $ are lower than $ V_d(\theta) $.
To make the above descriptions clear, we summarize the proposed robust sidelobe synthesis algorithm in Algorithm \ref{factodsafrps}.

\begin{algorithm}[!t]
	\caption{Proposed Robust Sidelobe Synthesis Algorithm}\label{factodsafrps}
	\begin{algorithmic}[1]
		\State give $ \theta_0 $, $ \varepsilon(\theta) $, the desired magnitude upper pattern $ V_d(\theta) $,
		the initial weight vector $ {\bf w}_{0} $ and its corresponding worst-case upper-boundary
		pattern $ V_{u,0}(\theta) $, set $ k=1 $
		\While{1}		
		\State select an angle $ \theta_k $ by comparing $ V_{u,k-1}(\theta) $ with
		$ ~~~~~V_d(\theta) $
		\State apply robust $ \textrm{C}^2\textrm{-WORD} $ to
		realize $V_{u,k}(\theta_k)=V_d(\theta_k) $, $ ~~~~ $
		obtain $ {\bf w}_{k} $ in \eqref{p21015} and the corresponding $ V_{u,k}(\theta) $
		\If{$ V_{u,k}(\theta){\rm ~is~not~satisfactory} $}
		\State set $ k=k+1 $
		\Else
		\State break
		\EndIf
		\EndWhile
		\State output $ {\bf w}_{k} $	
	\end{algorithmic}
\end{algorithm}

\begin{table}[!t]
	\renewcommand{\arraystretch}{1.15}
	\caption{Element Positions of The Non-uniform Linear Array}
	\label{table11}
	\centering
	\begin{tabular}{c | c ||c | c || c | c || c | c}
		\hline
		$~n~$&$x_n(\lambda)$&$~n~$&$x_n(\lambda)$&$~n~$&$x_n(\lambda)$&$~n~$&$x_n(\lambda)$\\
		\hline
		1&0.00 &4&1.55 &7&3.05 &10&4.55 \\
		2&0.45 &5&2.10 &8&3.65 &11&5.05 \\
		3&1.00 &6&2.60 &9&4.10 &12&5.50 \\								
		\hline
	\end{tabular}
\end{table}

\begin{figure*}[!t]
	\centering
	\subfloat[]
	{\includegraphics[width=3.0in]{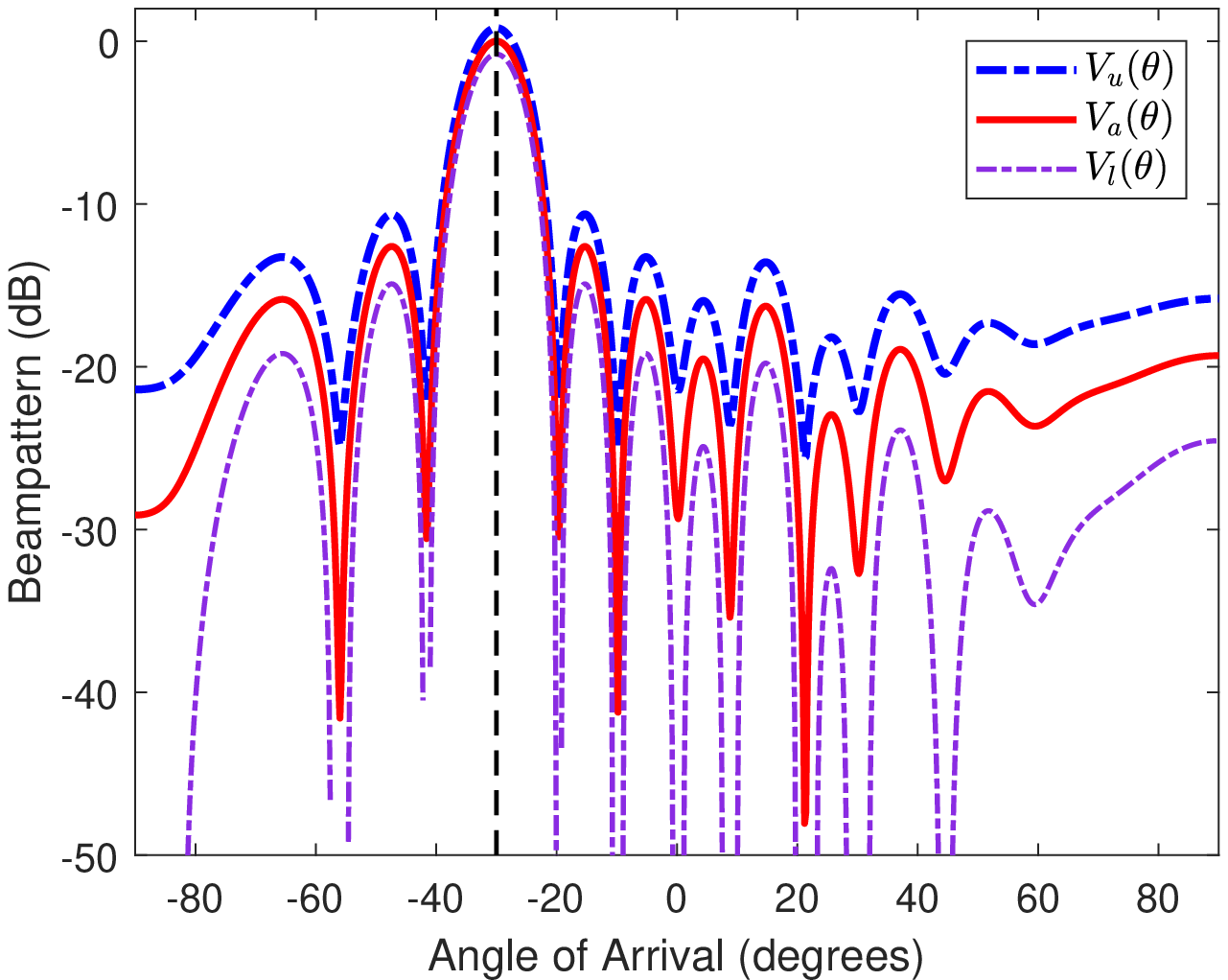}%
		\label{tspcontrolsrtep1allv3}}
	\hfil
	\subfloat[]
	{\includegraphics[width=3.0in]{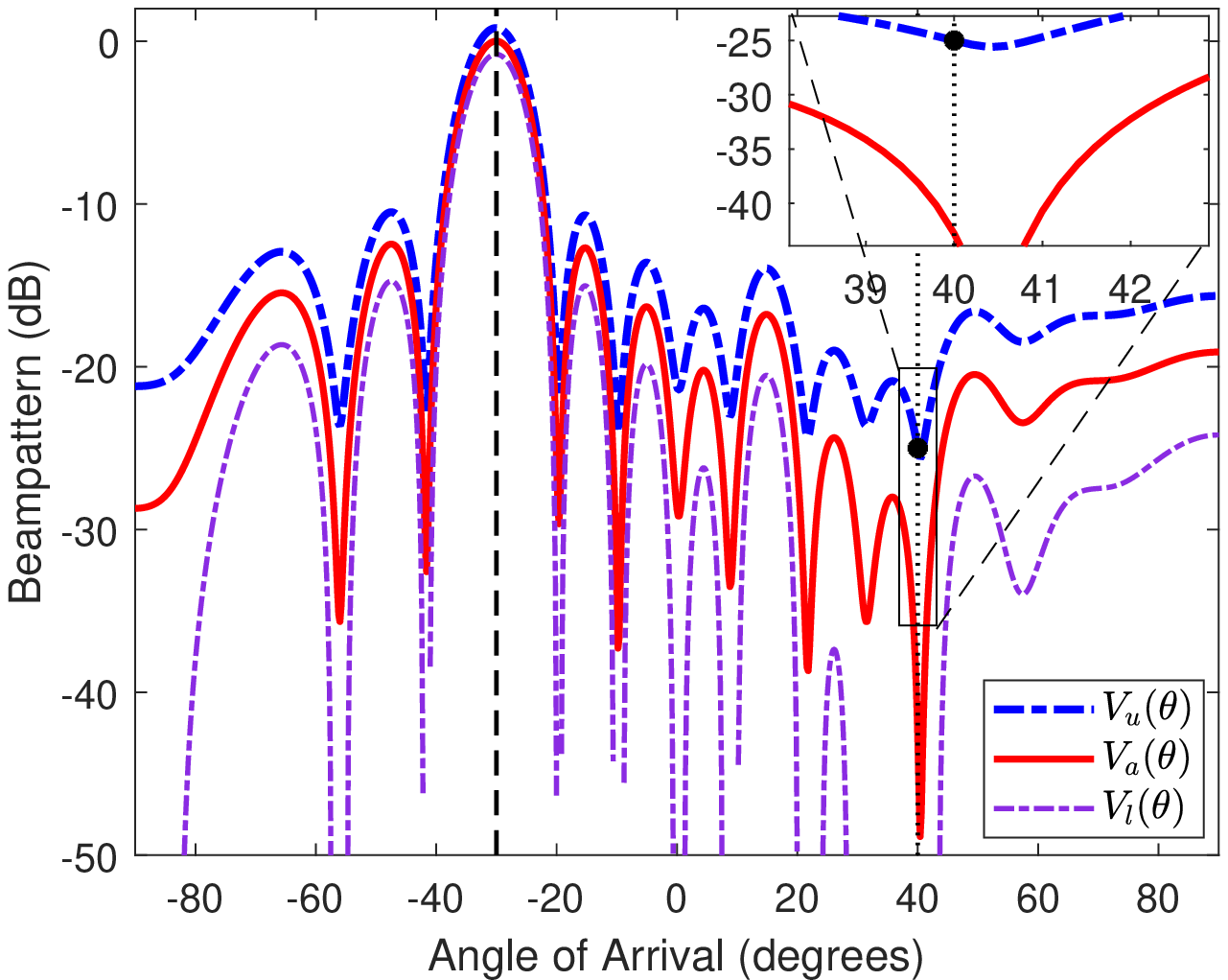}%
		\label{tspcontrrolstep2allv3}}\\							
	\caption{Illustration of response control for a non-uniformly spaced linear array.
		(a) Before array response control.
		(b) After array response control.}
	\label{tspcontrrolst}
\end{figure*}

\begin{figure*}[!t]
	\centering
	\subfloat[]
	{\includegraphics[width=3.0in]{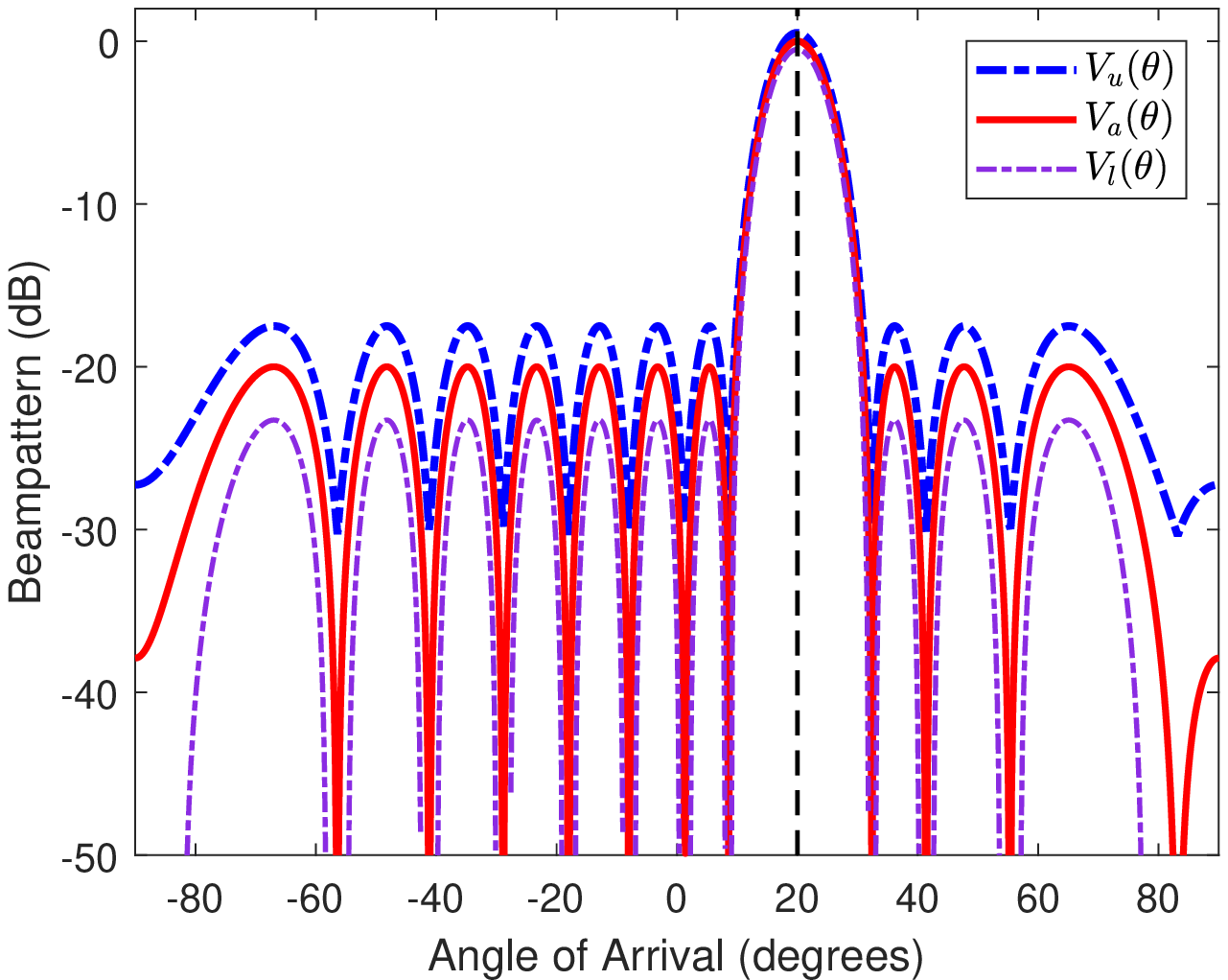}%
		\label{tspconrtrolstep1allv3}}
	\hfil
	\subfloat[]
	{\includegraphics[width=3.0in]{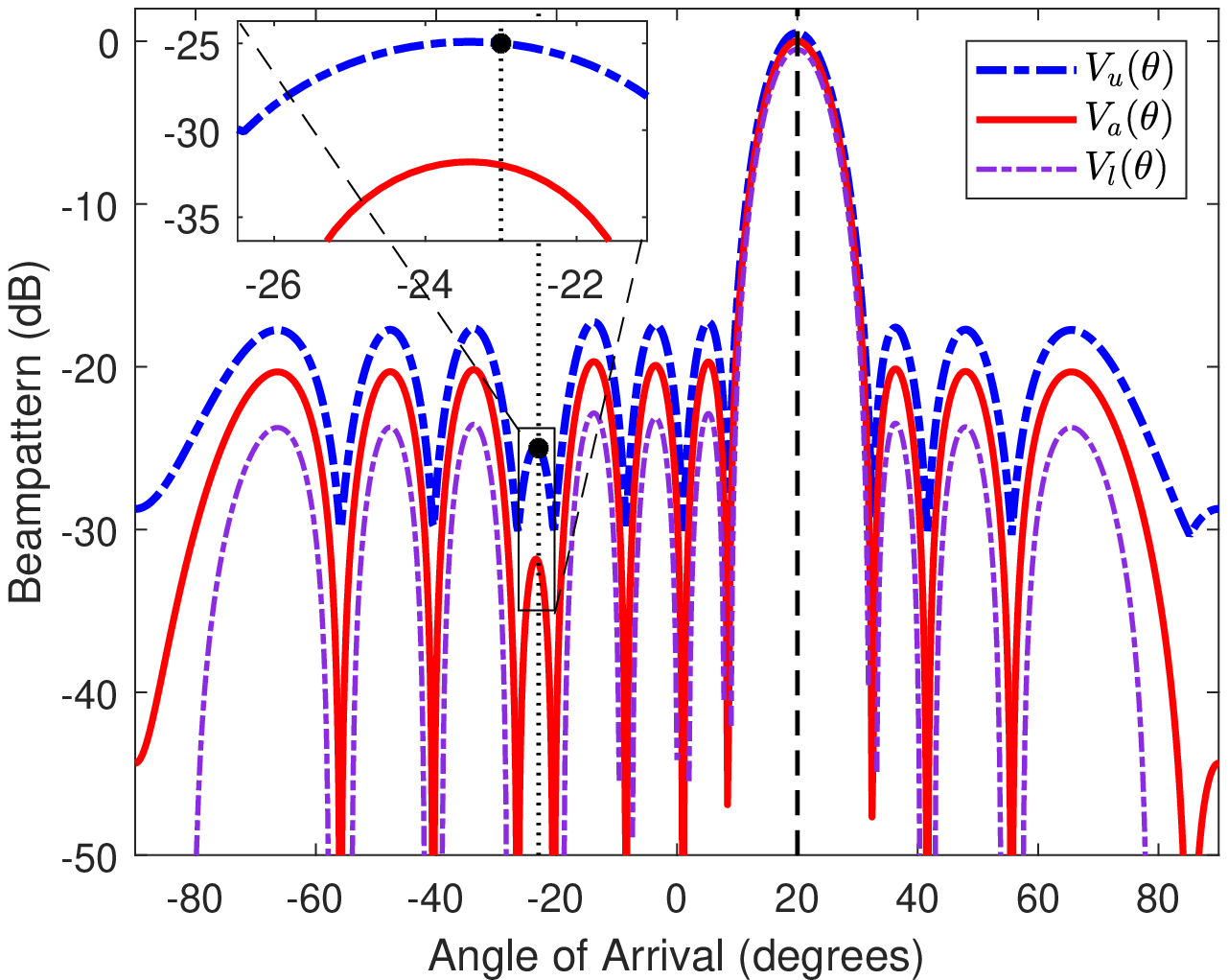}%
		\label{tsspcontrolstep2allv3}}\\							
	\caption{Illustration of response control for a uniformly spaced linear array.
		(a) Before array response control.
		(b) After array response control.}
	\label{tspcontrsrolst}
\end{figure*}

\section{Numerical Results}
We next present some simulations to show the effectiveness
of the proposed robust $ \textrm{C}^2\textrm{-WORD} $ scheme on array response
control and pattern synthesis under various settings.
To see the superiority of our algorithm, the results of
WORD algorithm in \cite{word} and
convex programming (CP) method in \cite{cwordref2} are also presented if 
applicable.
Unless otherwise specified, 
we take $ {\bf a}(\theta_0) $ as the initial weight for
both WORD and robust $ \textrm{C}^2\textrm{-WORD} $.

\subsection{Illustration of Robust Sidelobe Control}
In this part, we illustrate
the performance of robust $ \textrm{C}^2\textrm{-WORD} $ on
sidelobe control at a given point.
In the first example, we consider a non-uniformly spaced linear array with 12 elements, 
see Table \ref{table11} for its element positions.
The beam axis is steered to $ \theta_0=-30^{\circ} $ and the norm
boundary of steering vector is taken as $ \varepsilon(\theta)=0.16 $.
In this case, it is required to adjust the (actual) sidelobe response at $ \theta_1=40^{\circ} $ to
be lower than $ V_d(\theta_1)=-25{\rm dB} $.
Fig. \ref{tspcontrolsrtep1allv3} depicts the 
ideal beampattern $ V_a(\theta) $ of the initial weight vector $ {\bf w}_0={\bf a}(\theta_0) $,
the corresponding worst-case upper-boundary pattern $ V_u(\theta) $
and the lower-boundary pattern $ V_l(\theta) $.

Applying our robust $ \textrm{C}^2\textrm{-WORD} $ algorithm
and after some calculation, we can figure out that
$ {\rho}_{a,\star}=-42.7746{\rm dB} $, $ {\beta}_{1,\star}=0.077 $.
Fig. \ref{tspcontrrolstep2allv3} presents the resulting beampatterns
of the weight vector $ {\bf w}_1 $. It is shown that the upper-boundary
response level at $ {\theta_1} $ (i.e., $ V_u(\theta_1) $) has been
precisely adjusted to be $ V_d(\theta_1)=-25{\rm dB} $.
Since $ V_u(\theta) $ is the worst-case upper boundary of the sidelobe response,
we know that all the actual response level at $ \theta_1 $ is
lower than $ u_1 $, if only
$ \|\Delta(\theta)\|_2\leq\varepsilon(\theta) $ is satisfied.
In addition, comparing to the beampatterns in
Fig. \ref{tspcontrolsrtep1allv3}, it should be noted that the resulting beampatterns 
in Fig. \ref{tspcontrrolstep2allv3} are almost
unchanged at the uncontrolled points (i.e., $ \theta\neq\theta_1 $).
This benefit comes essentially from the advantage of $ \textrm{C}^2\textrm{-WORD} $
algorithm developed in Part I \cite{robust1}.

\begin{figure*}[!t]
	\centering
	\subfloat[]
	{\includegraphics[width=2.35in]{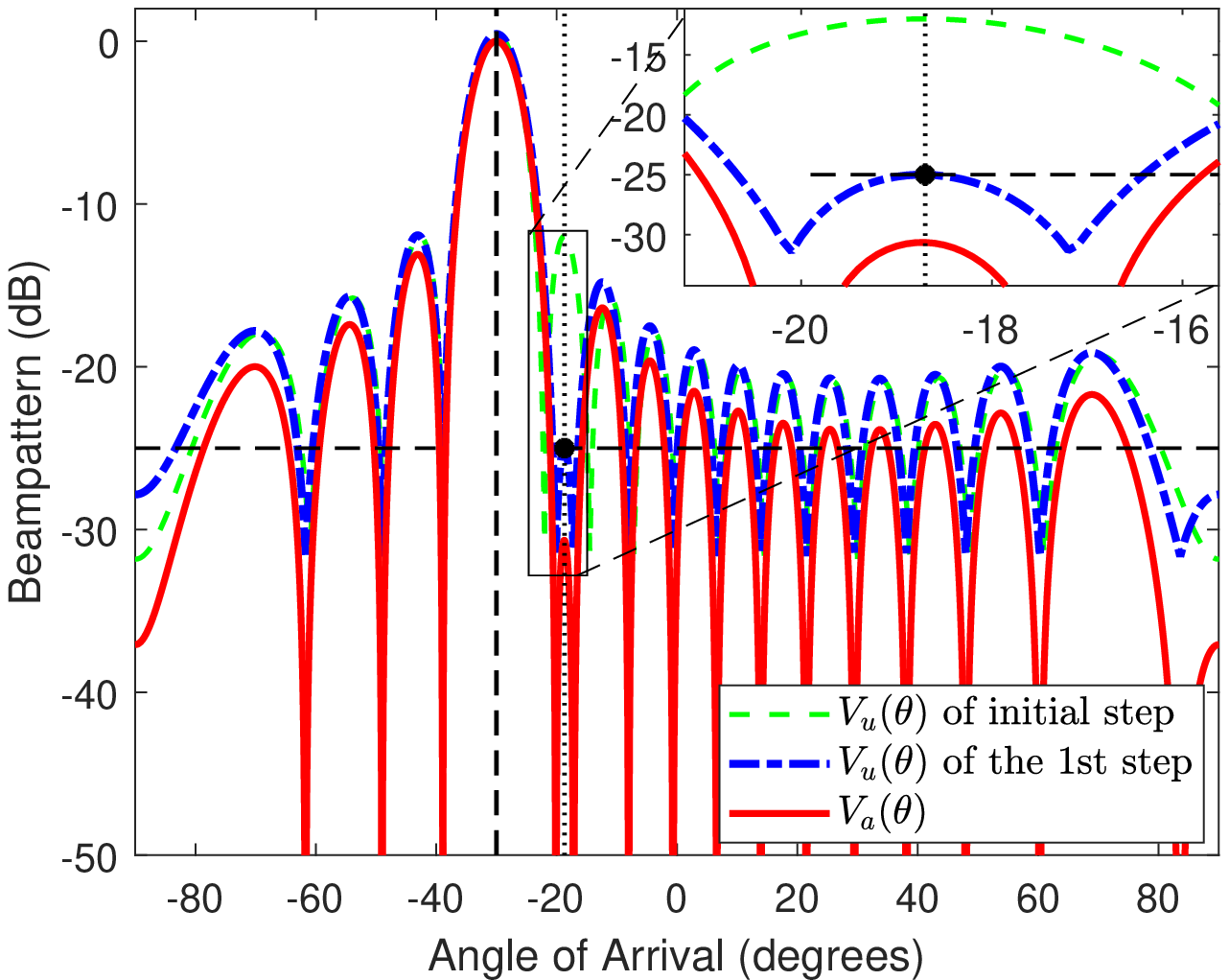}%
		\label{tapUrobustLAstep1}}
	\hfil
	\subfloat[]
	{\includegraphics[width=2.35in]{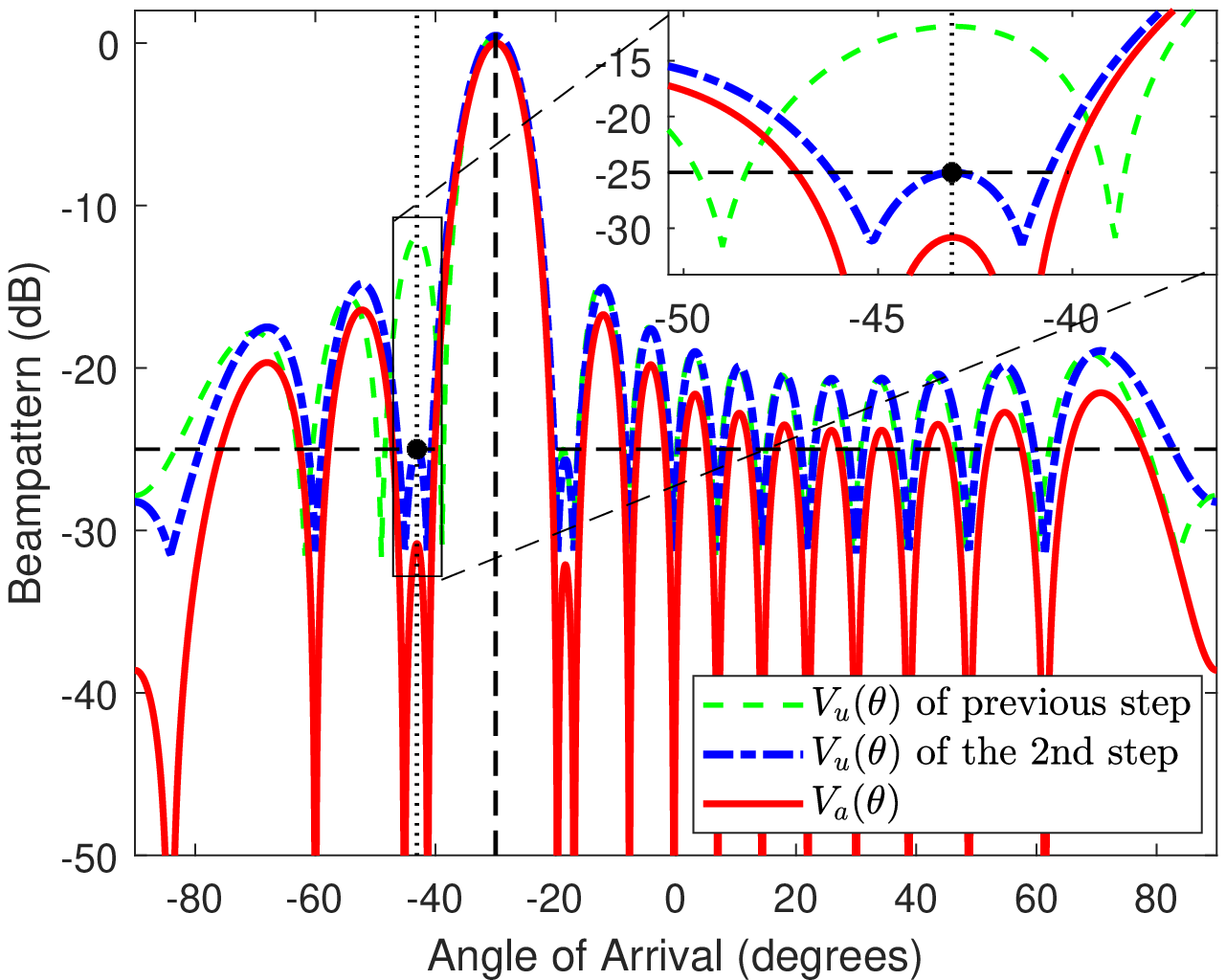}%
		\label{taprobustULAstep2}}	
	\subfloat[]
	{\includegraphics[width=2.35in]{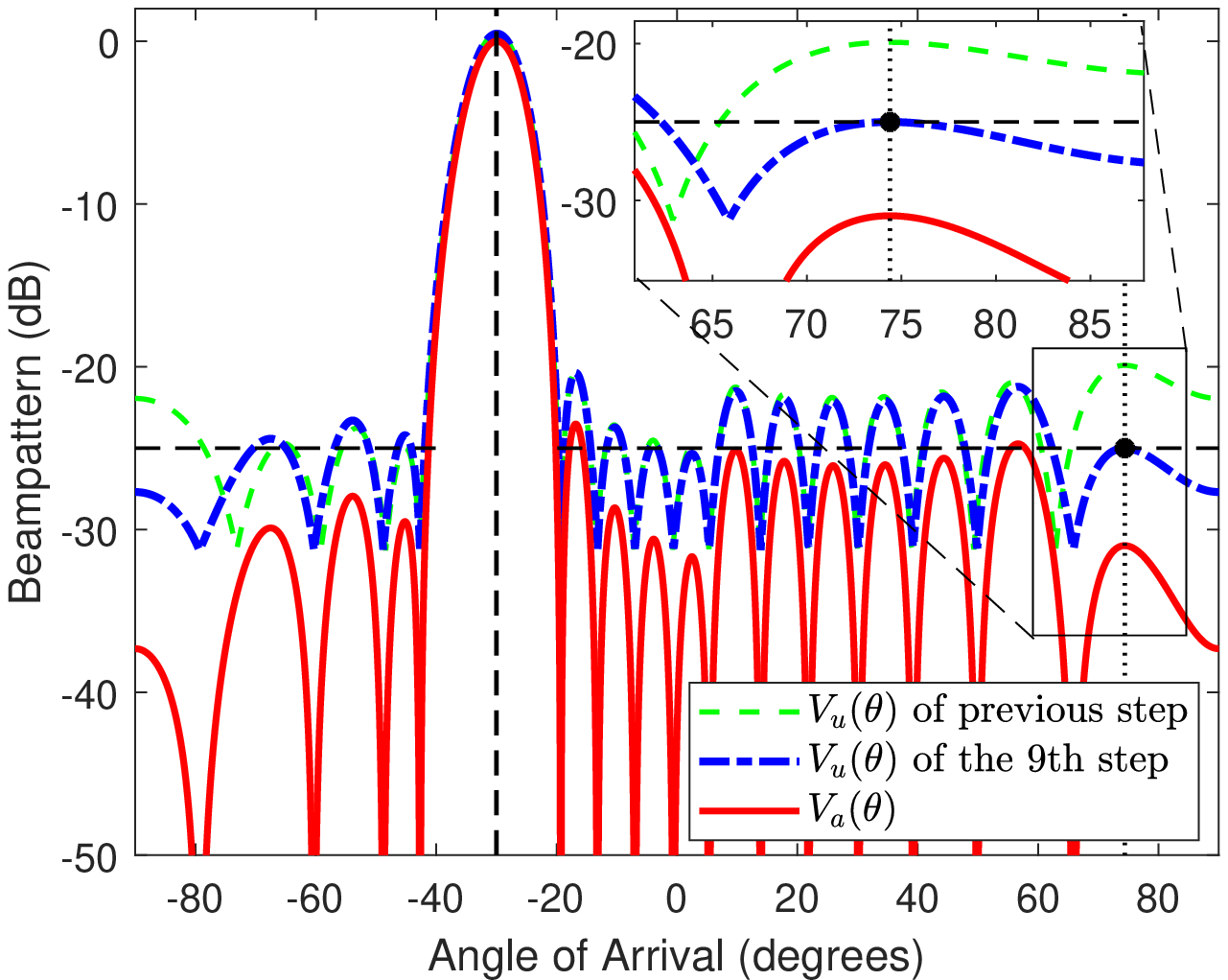}%
		\label{tapUrobustLAstep10}}\\							
	\caption{Synthesis procedure of uniform-sidelobe pattern using a ULA.
		(a) Synthesized pattern at the first step.
		(b) Synthesized pattern at the second step.
		(c) Synthesized pattern at the 9-th step.}
	\label{p2robustbf4}
\end{figure*}

To further show that our algorithm is effective for an arbitrarily-specified
initial weight, we consider a
12-element uniformly
spaced linear array (ULA) and steer its beam axis to $ \theta_0=20^{\circ} $.
In this case, we take the
initial weight of robust $ \textrm{C}^2\textrm{-WORD} $ as the Chebyshev weight with a 
$ -20{\rm dB} $ sidelobe attenuation.
The norm boundary of steering vector uncertainties is 
taken as $ \varepsilon(\theta)=0.1 $.
It is required to adjust the (actual) response level at $ \theta_1=-23^{\circ} $
to be lower than $ V_d(\theta_1)=-25{\rm dB} $.
Fig. \ref{tspconrtrolstep1allv3} shows the corresponding
$ V_a(\theta) $, $ V_u(\theta) $ and $ V_l(\theta) $ of the initial weight.
After carrying out the proposed robust $ \textrm{C}^2\textrm{-WORD} $ scheme,
we obtain that $ {\rho}_{a,\star}=-31.9987{\rm dB} $
and $ \beta_{1,\star}=0.2506 $.
The resulting beampatterns are presented in Fig. \ref{tsspcontrolstep2allv3},
from which we find that the value of $ V_u(\theta_1) $ equals exactly to 
$ V_d(\theta_1)=-25{\rm dB} $.
Also, it can be checked from Fig. \ref{tspcontrsrolst} that
our algorithm results small pattern variations
at the unadjusted points after the robust response control step.
In addition, we note that the resulting $ \beta_{1,\star} $'s are real-valued
in the above two testings. In fact, these are two special cases
that have been discussed in Part I \cite{robust1}.
A complex-valued $ \beta_{\star} $
will be resulted in general circumstances.

\subsection{Robust Sidelobe Synthesis Using Robust $ \textrm{C}^2\textrm{-WORD} $}
In this section, representative simulations are presented to
illustrate the application of 
robust $ \textrm{C}^2\textrm{-WORD} $ to sidelobe synthesis with
steering vector uncertainties.
\subsubsection{Uniform Sidelobe Synthesis for a ULA} 
In the first example, a 16-element ULA is considered. We steer the beam axis to $ \theta_0=-30^{\circ} $.
The desired upper beampattern has $ -25{\rm dB} $ uniform sidelobe level and
the steering vector uncertainty $ \Delta(\theta) $ is assumed to be norm-bounded by 
$ \varepsilon(\theta)=0.1 $.

Fig. \ref{p2robustbf4} presents several intermediate results when synthesizing pattern with 
robust $ \textrm{C}^2\textrm{-WORD} $ algorithm.
In the first step, our algorithm compares the worst-case upper-boundary pattern $ V_{u,0}(\theta) $ of the initial weight with the desired pattern $ V_d(\theta) $.
Then, we choose $ \theta_1=-18.7^{\circ} $ 
according to our angle selection strategy described in Section IV.
Applying robust $ \textrm{C}^2\textrm{-WORD} $, we figure
out that $ \rho_{a,\star}=-30.6544{\rm dB} $ and $ \beta_{1,\star}=0.1276 $
in the first step.
The resulting beampatterns are depicted in Fig. \ref{tapUrobustLAstep1},
from which we can see that the upper-boundary response level 
$ V_{u,1}(\theta_1) $
has been precisely adjusted as its desired level (i.e., $ -25{\rm dB} $).
Based on the resulting $ {\bf w}_1 $ and
$ V_{u,1}(\theta) $, we conduct the second step of robust $ \textrm{C}^2\textrm{-WORD} $
algorithm
and figure out that
$ \theta_2=-43.1^{\circ} $,
$ \rho_{a,\star}=-30.8132{\rm dB} $ and $ \beta_{1,\star}=0.1245 $.
The obtained beampatterns are illustrated in Fig. \ref{taprobustULAstep2}, from which 
we can check that $ V_{u,2}(\theta_2)=-25{\rm dB} $.
After applying the robust $ \textrm{C}^2\textrm{-WORD} $
algorithm iteratively, the envelope of $ V_u(\theta) $ becomes
closer to $ V_d(\theta) $, and we can terminate the iteration
if a satisfactory $ V_u(\theta) $ has been synthesized.

\begin{table}[!t]
	\renewcommand{\arraystretch}{1.15}
	\caption{Obtained Weightings of Robust $ \textrm{C}^2\textrm{-WORD} $ When 
		Synthesizing Uniform Sidelobe Pattern for a ULA}
	\label{tabl33e12}
	\centering
	\begin{tabular}{c | c ||c | c || c | c}
		\hline
		$~n~$&$ w_n $&$~n~$&$ w_n $&$~n~$&$ w_n $\\
		\hline
		1& $ 0.34e^{-j0.80} $ &7 & $ 1.25e^{+j2.35} $ &13&$ 0.77e^{-j0.78} $ \\
		2& $ 0.39e^{-j2.37} $ &8 & $ 1.31e^{+j0.78} $ &14&$ 0.58e^{-j2.35} $ \\
		3& $ 0.58e^{+j2.35} $ &9 & $ 1.31e^{-j0.78} $  &15&$ 0.39e^{+j2.37} $ \\
		4& $ 0.77e^{+j0.78} $ &10&$ 1.25e^{-j2.35} $  &16&$ 0.34e^{+j0.80} $ \\
		5& $ 0.96e^{-j0.79} $ &11&$ 1.13e^{+j2.36} $  & &  \\
		6& $ 1.13e^{-j2.36} $ &12&$ 0.96e^{+j0.79} $  & & \\									
		\hline
	\end{tabular}
\end{table}

\begin{figure}[!t]
	\centering
	\includegraphics[width=3.0in]{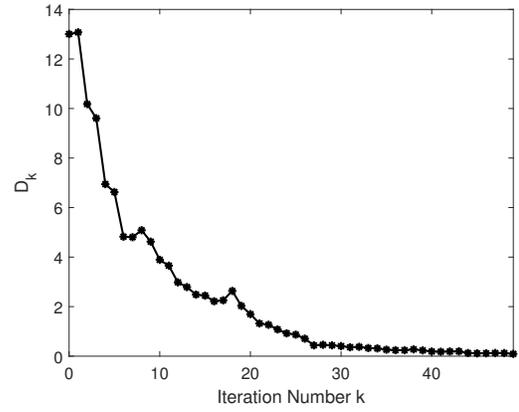}
	\caption{Maximum response deviation $ D_k $ versus the iteration number.}
	\label{robustdelta}
\end{figure}

\begin{figure}[!t]
	\centering
	\includegraphics[width=3.0in]{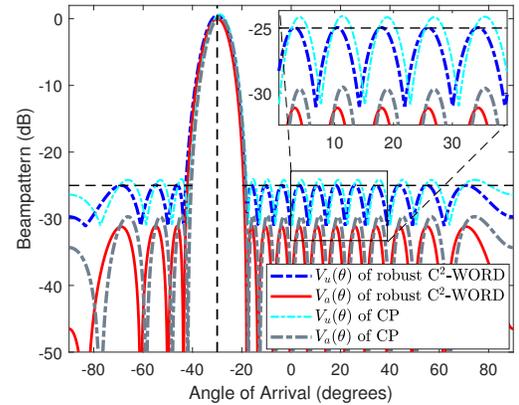}
	\caption{Synthesized patterns with uniform sidelobe for a ULA.}
	\label{uniformcompare}
\end{figure}

Following \cite{robust1}, we define $ D_k $ to
explore the convergence of the proposed approach.
More specifically, $ D_k $ measures the 
the maximum response deviation within the set of sidelobe peak angles at the $k$th step (denoted by $ {\Omega}_s^{k} $)
and is given by
\begin{align}
D_k\triangleq
\max_{\theta\in{\Omega}_s^{k}}
\left(
V_{u,k}(\theta)-V_d(\theta)
\right).
\end{align}	
The curve of $ D_k $ versus the iterative number $ k $ is depicted in
Fig. \ref{robustdelta}, which clearly shows that $ D_k $ decreases
with the increase of iteration.
After carrying out 50 response control steps, the resulting $ D_k $ 
equals approximately to zero and we terminate the synthesis process.
Table \ref{tabl33e12} presents the resulting weight vector of our
algorithm.
Interestingly, it is found that the weights are centro-symmetric. A possible explanation is that the array utilized has a symmetry structure.

The ultimate beampatterns are depicted in Fig. \ref{uniformcompare}.
We can see that the resulting worst-case upper-boundary pattern
$ V_u(\theta) $
of our algorithm aligns with the desired sidelobe level.
For the CP method in \cite{cwordref2}, we set the mainlobe region as $ [-42^{\circ},-18^{\circ}] $ and obtain an upper-boundary pattern
$ V_u(\theta) $ with a uniform sidelobe level, as shown in Fig. \ref{uniformcompare}.
For CP approach, the resulting maximum sidelobe level of $ V_u(\theta) $ is about $ -24{\rm dB} $, which is higher than that of 
the proposed robust $ \textrm{C}^2\textrm{-WORD} $ algorithm.
In fact, there is a trade-off between the mainlobe width
and the sidelobe level for CP method.
Moreover, it's not clear how to determine the mainlobe width
of CP approach
for a given sidelobe level requirement.

\begin{figure}[!t]
	\centering
	\includegraphics[width=3.0in]{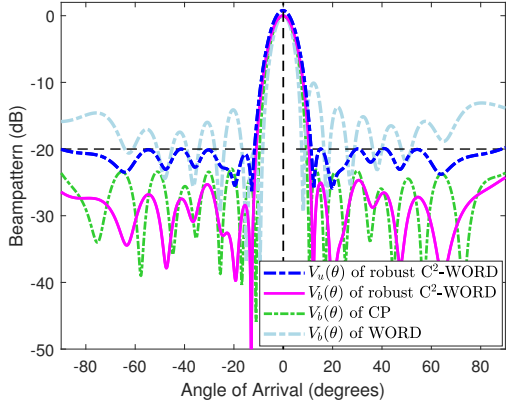}
	\caption{Synthesized patterns for a circular arc array 
	with gain-phase mismatch.}
	\label{phasegain}
\end{figure}

\begin{table}[!t]
	\renewcommand{\arraystretch}{1.15}
	\caption{Obtained Weightings of Robust $ \textrm{C}^2\textrm{-WORD} $ With Channel Phase-Gain Mismatch}
	\label{table12}
	\centering
	\begin{tabular}{c | c ||c | c || c | c}
		\hline
		$~n~$&$ w_n $&$~n~$&$ w_n $&$~n~$&$ w_n $\\
		\hline
		1& $ 0.27e^{-j1.56} $ &7 & $ 1.26e^{+j3.10} $ &13&$ 0.49e^{-j0.69} $ \\
		2& $ 0.37e^{+j1.42} $ &8 & $ 1.37e^{-j2.69} $ &14&$ 0.44e^{-j2.21} $ \\
		3& $ 0.43e^{-j2.16} $ &9 & $ 1.35e^{-j2.71} $  &15&$ 0.34e^{+j1.36} $ \\
		4& $ 0.50e^{-j0.60} $ &10&$ 1.25e^{+j3.06} $  &16&$ 0.26e^{-j1.67} $ \\
		5& $ 0.91e^{+j0.97} $ &11&$ 1.07e^{+j2.26} $  & &  \\
		6& $ 1.08e^{+j2.31} $ &12&$ 0.91e^{+j0.93} $  & & \\									
		\hline
	\end{tabular}
\end{table}

\subsubsection{Uniform Sidelobe Synthesis With Channel Phase-Gain Mismatch}
In this example, we consider a 
circular arc array with 16 nonisotropic elements, see 
Fig. 9 in \cite{robust1} with $ \theta_c=60^{\circ} $.
The beam axis is taken as $ \theta_0=0^{\circ} $
and the sidelobe level is expected to be lower than $ -20{\rm dB} $.
The distance between adjacent elements is
half a wavelength and there exists channel phase-gain
uncertainties on sensor elements.
More specifically, the phase error $ \varphi_n $
and gain error $ g_n $ are uniformly
distributed in $ [-0.035,0.035] $
and $ [0.98,1.02] $, respectively, $ n=2,\cdots,N $.
Following the analysis in Section III.A, we can
figure out $ \delta_1=0.039 $ and then obtain the
upper norm boundary $ \varepsilon(\theta) $ according to 
\eqref{eqn0998}.

Fig. \ref{phasegain} presents the resulting worst-case
upper-boundary pattern $ V_u(\theta) $ of
robust $ \textrm{C}^2\textrm{-WORD} $ algorithm after 20
iteration steps, and Table \ref{table12} lists the obtained weight vector.
It can be clearly observed that the sidelobe envelope 
of $ V_u(\theta) $ is aligned with the desired upper pattern 
$ V_d(\theta) $.
To compare the performances of different approaches,
Fig. \ref{phasegain} also demonstrates the
realizations of actual beampattern $ V_b(\theta) $.
We can see that the actual beampatterns of 
robust $ \textrm{C}^2\textrm{-WORD} $ and CP 
satisfy the pre-assigned response requirement,
while the WORD algorithm does not.

\begin{figure}[!t]
	\centering
	\includegraphics[width=3.0in]{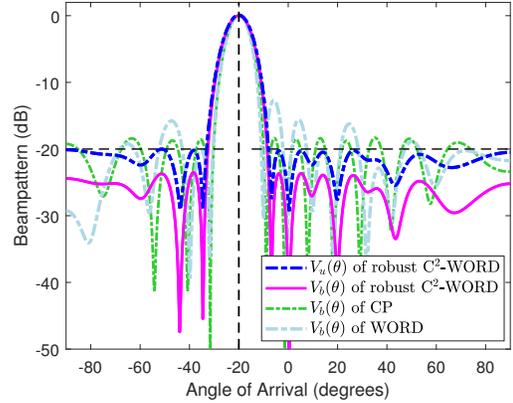}
	\caption{Synthesized patterns for a linear array 
		with element position mismatch.}
	\label{robustposition}
\end{figure}
\begin{table}[!t]
	\renewcommand{\arraystretch}{1.15}
	\caption{Obtained Weightings of Robust $ \textrm{C}^2\textrm{-WORD} $ With 
		Element Position Mismatch}
	\label{tab56le12}
	\centering
	\begin{tabular}{c | c ||c | c || c | c}
		\hline
		$~n~$&$ w_n $&$~n~$&$ w_n $&$~n~$&$ w_n $\\
		\hline
		1& $ 0.32e^{-j0.16} $ &5 & $ 1.08e^{+j1.78} $ &9&$ 0.77e^{-j2.60} $ \\
		2& $ 0.67e^{-j0.89} $ &6 & $ 1.17e^{+j0.77} $ &10&$ 0.64e^{+j2.73} $ \\
		3& $ 0.75e^{-j2.17} $ &7 & $ 1.07e^{-j0.40} $ &11&$ 0.48e^{+j1.80} $ \\
		4& $ 1.04e^{+j2.90} $ &8 &$ 1.08e^{-j1.47} $  &12&$ 0.34e^{+j0.62} $ \\									
		\hline
	\end{tabular}
\end{table}

\subsubsection{Uniform Sidelobe Synthesis With Element Position Mismatch}
We now carry out robust uniform sidelobe synthesis 
by considering array element position mismatch.
More specifically, we use a 12-element non-uniformly spaced
linear array, see Table \ref{table11} for its (ideal) element
positions.
The beam axis is steered to $ \theta_0=-20^{\circ} $
and the upper level of the desired sidelobe response 
is $ -20{\rm dB} $.
The array suffers from element position perturbation
and the location deviation is uniformly 
distributed in $ [-0.5\%\lambda,0.5\%\lambda] $.
Under these settings, we can determine $ \varepsilon(\theta) $
according to \eqref{epsi02} and realize robust sidelobe synthesis 
using our robust $ \textrm{C}^2\textrm{-WORD} $ algorithm.

Fig. \ref{robustposition} presents the resulting
$ V_u(\theta) $ of robust $ \textrm{C}^2\textrm{-WORD} $ algorithm
after 50 iteration steps, and Table \ref{tab56le12} gives
the obtained weight vector.
As expected, the obtained upper-boundary beampattern $ V_u(\theta) $
satisfies the pre-assigned response requirement.
To show the superiority of our algorithm,
we also depict the realizations of real beampattern $ V_b(\theta) $
for different methods.
Fig. \ref{robustposition} shows that
the real beampatterns of CP and WORD result
unqualified responses on sidelobe region.
For the robust $ \textrm{C}^2\textrm{-WORD} $ algorithm,
the obtained  $ V_b(\theta) $ meets our requirement with
a sidelobe level about $ -25{\rm dB} $.

\begin{figure}[!t]
	\centering
	\includegraphics[width=3.0in]{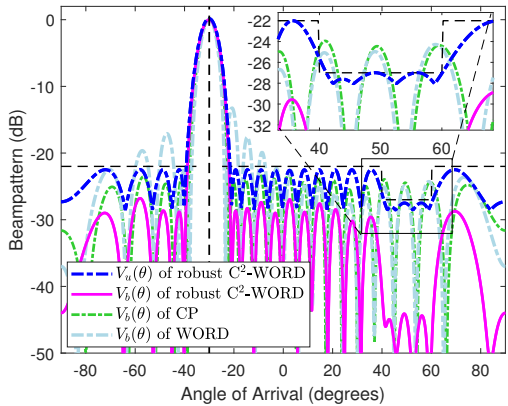}
	\caption{Synthesized patterns for a ULA
		with mutual coupling effect.}
	\label{mutual}
\end{figure}

\begin{table}[!t]
	\renewcommand{\arraystretch}{1.15}
	\caption{Obtained Weightings of Robust $ \textrm{C}^2\textrm{-WORD} $ With Mutual
	Coupling Effect}
	\label{tabdfle12}
	\centering
	\begin{tabular}{c | c ||c | c || c | c}
		\hline
		$~n~$&$ w_n $&$~n~$&$ w_n $&$~n~$&$ w_n $\\
		\hline
		1& $ 0.40e^{-j0.07} $ &8 & $ 1.16e^{+j1.58} $ &15&$ 0.95e^{+j3.14} $ \\
		2& $ 0.46e^{-j1.37} $ &9 & $ 1.28e^{+j0.01} $ &16&$ 0.84e^{+j1.55} $ \\
		3& $ 0.54e^{+j3.01} $ &10& $ 1.28e^{-j1.59} $ &17&$ 0.63e^{-j0.06} $ \\
		4& $ 0.63e^{+j1.64} $ &11&$ 1.28e^{-j3.12} $  &18&$ 0.54e^{-j1.44} $ \\
		5& $ 0.84e^{+j0.02} $ &12&$ 1.28e^{+j1.56} $  &19&$ 0.46e^{+j2.94} $ \\
		6& $ 0.95e^{-j1.57} $ &13&$ 1.16e^{-j0.01} $  &20&$ 0.40e^{+j1.65} $ \\
		7& $ 1.08e^{-j3.14} $ &14&$ 1.08e^{-j1.57} $  & & \\											
		\hline
	\end{tabular}
\end{table}

\subsubsection{Nonuniform Sidelobe Synthesis With Mutual Coupling Effect}
In this example, we consider a 20-element ULA
and set the beam axis as $ \theta_0=-30^{\circ} $.
Following the mutual coupling model in Section III.C,
we take mutual coupling effect into consideration
by setting the channel isolation as
$ \xi=-35{\rm dB} $.
With these configurations, one can readily
determine the upper boundary $ \varepsilon(\theta) $
from \eqref{upper3}.
Different from the previous testings,
in this case we consider a non-uniform desired upper sidelobe $ V_d(\theta) $.
More specifically, the upper level is
$ -22{\rm dB} $ in the region $ [40^{\circ},60^{\circ}] $ and 
$ -27{\rm dB} $ in the rest of the sidelobe region.

Fig. \ref{mutual} shows the resulting $ V_u(\theta) $ 
of the proposed robust $ \textrm{C}^2\textrm{-WORD} $ algorithm
after 80 iteration steps,
and Table \ref{tabdfle12} lists the corresponding weight vector.
Though the desired sidelobe level is non-uniform,
we can see clearly that $ V_u(\theta) $ 
satisfies the pre-assigned requirement.
Fig. \ref{mutual} also depicts the realizations
of real beampattern (i.e., $ V_b(\theta) $) for 
robust $ \textrm{C}^2\textrm{-WORD} $, CP and WORD.
It is observed that the robust $ \textrm{C}^2\textrm{-WORD} $
algorithm obtains a qualified beampattern $ V_b(\theta) $
with non-uniform sidelobe shape.
The resulting sidelobe of CP method is uniform and does not
satisfy the pre-assigned requirement in the null region.
As for WORD, the maximal sidelobe level
of $ V_b(\theta) $ is about $ -13{\rm dB} $, which is also an undesirable
result.

\newcounter{MYtempeqncnt}
\begin{figure*}[!t]
	\normalsize
	\setcounter{MYtempeqncnt}{\value{equation}}
	\setcounter{equation}{62}
	\begin{align}\label{y1y2}
	\left(
	\dfrac{{{\rho}_a|{\bf{w}}^{\mH}_{\bot}{\bf a}(\theta_0)|\cdot
			|{\bf{w}}^{\mH}_{\Arrowvert}{\bf a}(\theta_0)|}+
		\sqrt{{\rho}_a}|{\bf{w}}^{\mH}_{\bot}{\bf a}(\theta_0)|\cdot
		|{\bf{w}}^{\mH}_{\Arrowvert}{\bf a}(\theta_k)|}
	{\big|
		|{\bf{w}}^{\mH}_{\Arrowvert}{\bf a}(\theta_k)|^2-{\rho}_a
		|{\bf{w}}^{\mH}_{\Arrowvert}{\bf a}(\theta_0)|^2	
		\big|}
	\right)^2=
	\dfrac{
		\left(
		\dfrac{\sqrt{\rho_a}{\gamma(\theta_k)}}{V_d(\theta_k)-\sqrt{\rho_a}}
		\right)^2\|{\bf{w}}_{\bot}\|^2_2	
	}
	{
		|{\bf{w}}^{\mH}_{\Arrowvert}{\bf a}(\theta_k)|^2-\left(
		\dfrac{\sqrt{\rho_a}{\gamma(\theta_k)}}{V_d(\theta_k)-\sqrt{\rho_a}}
		\right)^2\|{\bf{w}}_{\Arrowvert}\|^2_2	
	}
	\end{align}	
	\hrulefill	
	\setcounter{equation}{\value{MYtempeqncnt}}
\end{figure*}
\begin{figure*}[!t]
	\normalsize
	\setcounter{MYtempeqncnt}{\value{equation}}
	\setcounter{equation}{64}
	\begin{subequations}\label{ABCDE}
		\begin{align}
		A&=(|{\bf{w}}^{\mH}_{\Arrowvert}{\bf a}(\theta_k)|^2-{\gamma^2(\theta_k)}
		\|{\bf{w}}_{\Arrowvert}\|^2_2)\cdot
		|{\bf{w}}^{\mH}_{\bot}{\bf a}(\theta_0)|^2\cdot|{\bf{w}}^{\mH}_{\Arrowvert}{\bf a}(\theta_0)|^2-
		{\gamma^2(\theta_k)}\|{\bf{w}}_{\bot}\|^2_2\cdot
		|{\bf{w}}^{\mH}_{\Arrowvert}{\bf a}(\theta_0)|^4\\
		B&=2\left(
		(|{\bf{w}}^{\mH}_{\Arrowvert}{\bf a}(\theta_k)|^2-{\gamma^2(\theta_k)}
		\|{\bf{w}}_{\Arrowvert}\|^2_2)
		-V_d(\theta_k)\cdot|{\bf{w}}^{\mH}_{\Arrowvert}{\bf a}(\theta_0)|
		|{\bf{w}}^{\mH}_{\Arrowvert}{\bf a}(\theta_k)|
		\right)
		|{\bf{w}}^{\mH}_{\bot}{\bf a}(\theta_0)|^2\cdot
		|{\bf{w}}^{\mH}_{\Arrowvert}{\bf a}(\theta_0)|\cdot
		|{\bf{w}}^{\mH}_{\Arrowvert}{\bf a}(\theta_k)|\\
		C&=V^2_d(\theta_k)|{\bf{w}}^{\mH}_{\Arrowvert}{\bf a}(\theta_k)|^2\cdot
		|{\bf{w}}^{\mH}_{\bot}{\bf a}(\theta_0)|^2\cdot
		|{\bf{w}}^{\mH}_{\Arrowvert}{\bf a}(\theta_0)|^2+
		(|{\bf{w}}^{\mH}_{\Arrowvert}{\bf a}(\theta_k)|^2-{\gamma^2(\theta_k)}
		\|{\bf{w}}_{\Arrowvert}\|^2_2)\cdot
		|{\bf{w}}^{\mH}_{\Arrowvert}{\bf a}(\theta_k)|^2\cdot
		|{\bf{w}}^{\mH}_{\bot}{\bf a}(\theta_0)|^2\nonumber\\
		&~~~-4V_d(\theta_k)|{\bf{w}}^{\mH}_{\Arrowvert}{\bf a}(\theta_k)|^3\cdot
		|{\bf{w}}^{\mH}_{\Arrowvert}{\bf a}(\theta_0)|\cdot
		|{\bf{w}}^{\mH}_{\bot}{\bf a}(\theta_0)|^2+2
		{\gamma^2(\theta_k)}\|{\bf{w}}_{\bot}\|^2_2\cdot
		|{\bf{w}}^{\mH}_{\Arrowvert}{\bf a}(\theta_k)|^2\cdot
		|{\bf{w}}^{\mH}_{\Arrowvert}{\bf a}(\theta_0)|^2\\
		D&=2V_d(\theta_k)|{\bf{w}}^{\mH}_{\Arrowvert}{\bf a}(\theta_k)|^3\cdot
		|{\bf{w}}^{\mH}_{\bot}{\bf a}(\theta_0)|^2\cdot
		(
		V_d(\theta_k)|{\bf{w}}^{\mH}_{\Arrowvert}{\bf a}(\theta_0)|-
		|{\bf{w}}^{\mH}_{\Arrowvert}{\bf a}(\theta_k)|
		)\\
		E&=|{\bf{w}}^{\mH}_{\Arrowvert}{\bf a}(\theta_k)|^4\cdot
		\left(
		V^2_d(\theta_k)|{\bf{w}}^{\mH}_{\bot}{\bf a}(\theta_0)|^2-
		{\gamma^2(\theta_k)}\cdot
		\|{\bf{w}}_{\bot}\|^2_2
		\right)
		\end{align}
	\end{subequations}		
	\hrulefill	
	\setcounter{equation}{\value{MYtempeqncnt}}
\end{figure*}

\section{Conclusions}
In this paper, we have presented a new algorithm named robust
$ \textrm{C}^2\textrm{-WORD} $, which can realize robust sidelobe control and synthesis with steering vector mismatch.
The proposed robust $ \textrm{C}^2\textrm{-WORD} $ algorithm 
offers an analytical expression of weight vector updating
and is able
to precisely control the worst-case upper-boundary
response level of a given sidelobe point for
the norm-bounded steering vector uncertainties.
We have also presented detailed analyses on how to determine the
norm boundary of steering vector uncertainty 
under various mismatch circumstances.
Moreover, a robust sidelobe synthesis approach
has been devised by successively applying 
robust $ \textrm{C}^2\textrm{-WORD} $ algorithm.
The applications of robust $ \textrm{C}^2\textrm{-WORD} $
to robust sidelobe control and synthesis have been
validated with various examples.
As a future work, we shall consider
the robust multi-point response control 
algorithm so as to reduce the number of iteration
step in synthesis process.

\appendices


\section{Derivation of \eqref{qus}}
To simplify the notation, we omit the subscript of $ \beta $ 
in sequel.
According to the definition of $ \rho_a $ in \eqref{rhoa},
we have
\begin{align}\label{eqn01}
\sqrt{\rho_a}={|{\bf w}^{\mH}_{k}{\bf a}(\theta_k)|}/{|{\bf w}^{\mH}_{k}{\bf a}(\theta_0)|}.
\end{align}
Recalling the constraint \eqref{p21118}, we can expand it as
\begin{align}\label{eqn03}
\sqrt{\rho_a}
&=V_d(\theta_k)-\dfrac{\gamma(\theta_k)\|{\bf w}_{k}\|_2}
{|{\bf w}^{\mH}_{k}{\bf a}(\theta_0)|}\\
&=V_d(\theta_k)-\dfrac{\sqrt{\rho_a}{\gamma(\theta_k)}\|{\bf w}_{k}\|_2}
{|{\bf w}^{\mH}_{k}{\bf a}(\theta_k)|}
\end{align}
where Eqn. \eqref{eqn01} has been utilized.
Substituting the constraint \eqref{p211118} into $ {\bf w}_k $, one obtains
\begin{align}\label{eqn04}
\dfrac{|{\bf w}^{\mH}_k{\bf a}(\theta_k)|^2}
{\|{\bf w}_{k}\|^2_2}=\dfrac{|\beta|^2|{\bf{w}}^{\mH}_{\Arrowvert}{\bf a}(\theta_k)|^2}
{\|{\bf{w}}_{\bot}\|^2_2+|\beta|^2\|{\bf{w}}_{\Arrowvert}\|^2_2}.
\end{align}
Then, we can reformulate
Eqn. \eqref{eqn03} as
\begin{align}\label{eqn05}
V_d(\theta_k)-\sqrt{\rho_a}=\dfrac{\sqrt{\rho_a}{\gamma(\theta_k)}}
{\sqrt{\dfrac{|\beta|^2|{\bf{w}}^{\mH}_{\Arrowvert}{\bf a}(\theta_k)|^2}
		{\|{\bf{w}}_{\bot}\|^2_2+|\beta|^2\|{\bf{w}}_{\Arrowvert}\|^2_2}}}
\end{align}
or equivalently,
\begin{align}\label{eqn07}
|\beta|^2=\dfrac{
	\left(
	\dfrac{\sqrt{\rho_a}{\gamma(\theta_k)}}{V_d(\theta_k)-\sqrt{\rho_a}}
	\right)^2\|{\bf{w}}_{\bot}\|^2_2	
}
{
	|{\bf{w}}^{\mH}_{\Arrowvert}{\bf a}(\theta_k)|^2-\left(
	\dfrac{\sqrt{\rho_a}{\gamma(\theta_k)}}{V_d(\theta_k)-\sqrt{\rho_a}}
	\right)^2\|{\bf{w}}_{\Arrowvert}\|^2_2	
}.
\end{align}

On the other hand, recalling from Proposition 2 in Part I \cite{robust1} that $ {\beta} $ satisfies
\begin{align}\label{eqn08}
|{\beta}|=|{\bf c}_{\beta}|+R_{\beta}
\end{align}
with
\begin{align}\label{eqn09}
|{\bf c}_{\beta}|&=\dfrac{{\rho}_a|{\bf{w}}^{\mH}_{\bot}{\bf a}(\theta_0)|\cdot
	|{\bf{w}}^{\mH}_{\Arrowvert}{\bf a}(\theta_0)|}
{|{\bf B}(2,2)|}\\
R_{\beta}&=\dfrac{\sqrt{{\rho}_a}|{\bf{w}}^{\mH}_{\bot}{\bf a}(\theta_0)|\cdot
	|{\bf{w}}^{\mH}_{\Arrowvert}{\bf a}(\theta_k)|}
{|{\bf B}(2,2)|}
\end{align}
where $ {\bf B} $ is given by
\begin{align}
\label{close004}\!\!\!{\bf B}=\!\!\begin{bmatrix}
{\bf{w}}^{\mH}_{\bot}{\bf{a}}(\theta_k)\\ {\bf{w}}^{\mH}_{\Arrowvert}{\bf{a}}(\theta_k)
\end{bmatrix}\!\!
\begin{bmatrix}
{\bf{w}}^{\mH}_{\bot}{\bf{a}}(\theta_k)\\ {\bf{w}}^{\mH}_{\Arrowvert}{\bf{a}}(\theta_k)
\end{bmatrix}^{\mH}\!\!\!\!\!-\!
\rho_a\!\!
\begin{bmatrix}
{\bf{w}}^{\mH}_{\bot}{\bf{a}}(\theta_0)\\ {\bf{w}}^{\mH}_{\Arrowvert}{\bf{a}}(\theta_0)
\end{bmatrix}\!\!
\begin{bmatrix}
{\bf{w}}^{\mH}_{\bot}{\bf{a}}(\theta_0)\\ {\bf{w}}^{\mH}_{\Arrowvert}{\bf{a}}(\theta_0)
\end{bmatrix}^{\mH}\!\!\!\!.
\end{align}
Thus, we have
\begin{align}
|{\beta}|=\dfrac{{{\rho}_a|{\bf{w}}^{\mH}_{\bot}{\bf a}(\theta_0)|\cdot
		|{\bf{w}}^{\mH}_{\Arrowvert}{\bf a}(\theta_0)|}+
	\sqrt{{\rho}_a}|{\bf{w}}^{\mH}_{\bot}{\bf a}(\theta_0)|\cdot
	|{\bf{w}}^{\mH}_{\Arrowvert}{\bf a}(\theta_k)|}
{\big|
	|{\bf{w}}^{\mH}_{\Arrowvert}{\bf a}(\theta_k)|^2-{\rho}_a
	|{\bf{w}}^{\mH}_{\Arrowvert}{\bf a}(\theta_0)|^2	
	\big|}.\nonumber
\end{align}

Substituting the above $ |{\beta}| $ into \eqref{eqn07}, we
can eliminate $ \beta $ and obtain a quartic equation with respect to $ {\rho_a} $ 
as shown in \eqref{y1y2} on the top of this page.
After some manipulations, one can reshape \eqref{y1y2} as
\addtocounter{equation}{1}
\begin{align}\label{eqn11}
A{\rho}^2_k+B{\rho_a}\sqrt{\rho_a}+C\rho_a+D\sqrt{\rho_a}+E=0
\end{align}
where the coefficients $ A $, $ B $, $ C $, $ D $ and $ E $ are given in
\eqref{ABCDE} on the top of this page.
Eqn. \eqref{eqn11} can be alternatively expressed as
\addtocounter{equation}{1}
\begin{align}\label{eqn13}
A{\rho}^2_k+C\rho_a+E=-\sqrt{\rho_a}(B{\rho_a}+D).
\end{align}
Taking square to both sides of \eqref{eqn13} yields
\begin{align}
A^2{\rho}^4_a+(2AC-B^2){\rho}^3_a+(2AE-2BD+C^2){\rho}^2_a\nonumber\\
~~~~~~~+(2CE-D^2){\rho}_a+E^2=0.
\end{align}
This completes the derivation of \eqref{qus}.



\bibliography{C2WORD20718}

\end{document}